\documentclass[useAMS,usenatbib,fleqn]{mnras}
\usepackage{graphicx,color,amsmath,amssymb,amsfonts}
\usepackage{subfig}
\usepackage[toc,page]{appendix}
\usepackage{hyperref} 
\usepackage{etoolbox}
\usepackage{natbib}
\usepackage{rotating}

\makeatletter
\patchcmd\@combinedblfloats{\box\@outputbox}{\unvbox\@outputbox}{}{%
  \errmessage{\noexpand\@combinedblfloats could not be patched}%
}%
\makeatother

\newcommand{\be}{\begin{equation}}
\newcommand{\ee}{\end{equation}}
\newcommand{\bea}{\begin{eqnarray}}
\newcommand{\eea}{\end{eqnarray}}

\newcommand{\F}[1]{{Fig. \ref{#1}}}
\newcommand{\T}[1]{{Table \ref{#1}}}
\newcommand{\Eq}[1]{{Eq. \ref{#1}}}
\newcommand{\tb}[1]{{\boldsymbol{#1}}}
\newcommand{\nest}{\textsc{MultiNest}}
\newcommand{\me}{\text{e}}

\title[1.4-GHz quasar luminosity functions]{The optically-selected 1.4-GHz quasar luminosity function below 1\,mJy}
\author[Malefahlo ~et~al.]{Eliab Malefahlo\thanks{eliabmalefahlo3@gmail.com}$^1$, Mario G. Santos$^{1,2,3}$, Matt J.~Jarvis$^{4,1}$,  Sarah V. White$^{5,2}$, \and Jonathan T.~L.~Zwart$^{7,1,6}$\\\\
$^{1}$Department of Physics \& Astronomy, University of the Western
Cape, Private Bag X17, Bellville, Cape Town, 7535, South Africa\\
$^{2}$South African Radio Astronomy Observatory (SARAO), 2 Fir Street, Observatory, Cape Town, 7925, South Africa\\
$^{3}$Instituto de Astrofisica e Ciencias do Espaco, Universidade de Lisboa, OAL, Tapada da Ajuda, PT1349-018 Lisboa, Portugal\\
$^{4}$Astrophysics, Department of Physics, Keble Road, Oxford, OX1 3RH, UK\\
$^{5}$Department of Physics and Electronics, Rhodes University, PO Box 94, Grahamstown, 6140, South Africa\\
$^{6}$Department of Astronomy, University of Cape Town, Private Bag X3, Rondebosch, Cape Town, 7701, South Africa\\
$^{7}$SprintHive (Pty) Ltd, Brickfield Canvas, 35 Brickfield Road, Salt River, Cape Town, 7925, South Africa\\
}

\begin{document}

\date{Accepted 8 January 2020. Received 8 January 2020; in original form 15 August 2019}

\pagerange{\pageref{firstpage}--\pageref{lastpage}} \pubyear{2019}

\maketitle

\label{firstpage}
\begin{abstract}
\noindent
We present the radio luminosity function (RLF) of optically-selected quasars below 1~mJy, constructed by applying a Bayesian-fitting stacking technique to objects well below the nominal radio flux-density limit. We test the technique using simulated data, confirming that we can reconstruct the RLF over three orders of magnitude below the typical $5\sigma$ detection threshold. We apply our method to 1.4-GHz flux-densities from the Faint Images of the Radio Sky at Twenty-cm survey (FIRST), extracted at the positions of optical quasars from the Sloan Digital Sky Survey (SDSS) over seven redshift bins up to $z=2.15$ and measure the RLF down to two orders of magnitude below the FIRST detection threshold. In the lowest redshift bin ($0.2<z<0.45$), we find that our measured RLF agrees well with deeper data from the literature. The RLF for the radio-loud quasars flattens below $\log_{10}[L_{1.4}/{\rm W\,Hz}^{-1}] \approx 25.5$ and becomes steeper again below $\log_{10}[L_{1.4}/{\rm W\,Hz}^{-1}] \approx 24.8$, where radio-quiet quasars start to emerge. The radio luminosity where radio-quiet quasars emerge coincides with the luminosity where star-forming galaxies are expected to start to dominate the radio source counts. This implies that there could be a significant contribution from star formation in the host galaxies, but additional data is required to investigate this further. The higher-redshift bins show a similar behaviour as for the lowest-$z$ bin, implying that the same physical process may be responsible.

\end{abstract}


\section{Introduction}
The evolution of quasars has been a subject of interest right since their discovery \citep{Schmidt-1963}. Quasars have been of particular interest over the past decade due to the role that they --- and active galactic nuclei (AGN) in general --- play in galaxy evolution. For example, feedback from AGN may expel or heat gas in a galaxy, thereby quenching star formation (SF) in the host galaxies \citep[e.g.][]{Granato-2004,Scannapieco_Oh-2004,Croton-2006,Hopkins-2008,Antonuccio-Delogu_Silk-2008}, or feasibly in the wider environment \citep[e.g.][]{RawlingsJarvis-2004,Hatch-2014}. This may be a major contributor to establishing the observed relationship between supermassive black holes (SMBHs) and the central bulge properties in a galaxy (e.g.~\citealt{Ferrarese_Merritt-2000,Hopkins-2006}). 

They were originally discovered as strong radio sources and later also found to be bright in the optical \citep[e.g.][]{Schmidt-1963}. However only $\sim 10$~per~cent of optically-selected quasars were detected in large-area radio surveys (e.g. \citealt{Strittmatter-1980}). The sources that were detected in these surveys were termed `radio-loud quasars', while the remaining 90~per~cent of the quasar population, which are fainter in the radio, were referred to as `radio-quiet quasars'.
The radio emission from radio-loud quasars is known to be mainly dominated by synchrotron radiation from electrons accelerated by powerful jets, while the source of radio-quiet quasars is still debated. One suggestion is that the radio emission from radio-quiet quasars is a result of synchrotron radiation from supernova explosions associated with star formation in the host galaxy, rather than being the result of AGN processes (e.g.~\citealt{Terlevich-1987,Terlevich-1992,Padovani-2011,Kimball-2011,Bonzini-2013,Condon-2013,Kellermann-2016,Gurkan-2018,Stacey-2018}). However, some authors suggest the radio emission in radio-quiet quasars is still dominated by AGN-related processes such as low-power jets \citep[e.g.][]{FalckeBiermann-1995,Wilson_Colbert-1995,Hartley-2019}, accretion disk winds \citep[e.g.][]{Jiang-2010,Zakamska-2014}, coronal disk emissions \citep{LoarBehar-2008,Laor-2019} or a combination of these process  (see \citealt{Panessa-2019} for a review), with factors such as different accretion rates \citep{Fernandes-2011}, SMBH spin \citep{Blandford_Znajek-1977,Schuize-2017}, SMBH mass \citep{Dunlop-2003,McLureJarvis2004}, host-galaxy morphology \citep{Bessiere-2012}, galactic environments \citep{fan-2001a}, or a combination of these, being responsible for lack of powerful jets .
 
One of the ways to study quasars and their source of radio emission is through luminosity functions (LFs, i.e.~the number of sources with a certain luminosity in a given volume and luminosity bin). 
It is now accepted that SMBHs accrete most of their mass during the active-galaxy phase, when they are radiating at quasar luminosities \citep{Salpeter-1964,Zeldovich_Novikov-1965,Lynden_Bell-1969,Soltan-1982}. Therefore, with accurate measurements of the quasar LF and its evolution, one can map out the SMBH accretion history (e.g.~ \citealt{Shankar-2009,Shankar-2010,Shen-2009,Shen_Kelly-2012}), constrain the formation history of SMBHs (e.g.~\citealt{Rees-1984}, \citealt{Haiman-2012}), and potentially determine the contribution of quasars to feedback. 

The radio luminosity function (RLF) of radio-loud quasars is well-studied \citep[e.g.][]{Schmidt-1970,Willott-1998,Jiang-2007}, but the faint (radio-quiet) end is not well-explored, as these fainter sources lie below the detection threshold of most wide-area radio surveys. There are various methods used in the literature to study radio-quiet populations. One such method is through deep-narrow radio surveys \citep[e.g.][]{Condon-2003,Kellermann-2008, Padovani-2009, Padovani-2011, Miller-2013}. Such surveys have contributed to our understanding of the radio emission from the radio-quiet population. For instance, \citet{Padovani-2015} found that emission from radio-quiet AGNs has a contribution from black-hole activity as well as emission related to star formation. However, very few genuinely luminous quasars are detected in these deep-narrow survey ($\sim$~15 quasars per deg$^2$). 
 
The most popular means of studying $\mu$Jy sources in the past two decades have involved some form of `stacking' \citep{Ivezic-2002, White-2007, Hodge-2008, Mitchell_Wynne-2013, Roseboom_Best-2014, Zwart-2014a}. There are a number of different versions and definitions of stacking seen in the literature (see \citealt{Zwart-2014a} for an overview). Usually stacking involves using positional information of a source population that is selected (and classified) from an auxiliary survey, and then extracting the flux density at those positions in the survey of interest (where they are above or below the detection threshold). In most cases stacking is used to explore the average (mean, median or weighted versions thereof) properties of sources below the detection threshold (i.e. $<<5\sigma$). For example, stacking can be employed to infer average SF rates (e.g.~\citealt{Dunne-2009}; \citealt{Karim-2011}, \citealt{Zwart-2014}), where 1.4-GHz radio flux-densities are extracted at positions of sources selected by stellar mass.

Traditional stacking techniques have added a great deal to our understanding of $\mu$Jy source populations. However, they only return a single statistic, and new techniques have been developed that extract more information from the stacked data. \citet{Mitchell_Wynne-2013} went beyond stacking by combining stacking with maximum-likelihood methods to fit a source-count model to the stacked sources. \citet{Roseboom_Best-2014} adopted a similar approach to \citet{Mitchell_Wynne-2013} by fitting a luminosity-function model to stacked star-forming galaxies. \citet{Zwart-2015} then extended the technique of \citet{Mitchell_Wynne-2013} to a fully-Bayesian framework (\textsc{bayestack}), which allows for model selection.
\citet{Chen-2017} extended the technique of \citet{Zwart-2015} by including the effects of the point spread function and source confusion, an approach that incorporates some of the reasoning from \citet{Vernstrom-2014}, which combined a traditional $P(D)$ analysis with a Bayesian likelihood model fitting.

In this work we measure the RLF of optically-selected quasars below 1~mJy by building on the work of \cite{Roseboom_Best-2014} and \cite{Zwart-2015}. We use a set of models for the RLF and fit directly to the radio data using a full Bayesian approach.
We apply the technique to a large sample of quasars from the Sloan Digital Sky Survey (SDSS; \citealt{York-2000}) Data release 7 (DR7) quasar catalogue \citep{Shen-2011}, using flux densities taken from the Faint Images of The Radio Sky at Twenty-centimeters (FIRST; \citealt{Becker-1995}).

In Section~\ref{sec:data} we describe the optical and radio data used in this study. We then outline our technique for making measurements below the noise level using \textsc{bayestack} (Section~\ref{sec:Below}). In Section~\ref{sec:Test} we test the technique, and our results are given in Section~\ref{sec:results}. We discuss the results and compare them to the literature in Section~\ref{sec:discusion}, finally concluding in Section~\ref{sec:conclusion}.
Throughout the work, unless stated otherwise, we use AB magnitudes and the positions are in the J2000 equinox. We set the spectral index, defined as $\alpha \equiv \log(S/S_0)/ \log(\nu_0/\nu)$, to $\alpha=0.7$ \citep[e.g.][]{Kukula-1998} when converting flux density to luminosity and one reference frequency to another.  We assume a $\Lambda$CDM cosmology, with $H_0 = 70$ km$^{-1}$ Mpc$^{-1}$, $\Omega_\Lambda = 0.7$ and $\Omega_m = 0.3$.

\begin{figure}
\centering
\includegraphics[width=0.5\textwidth]{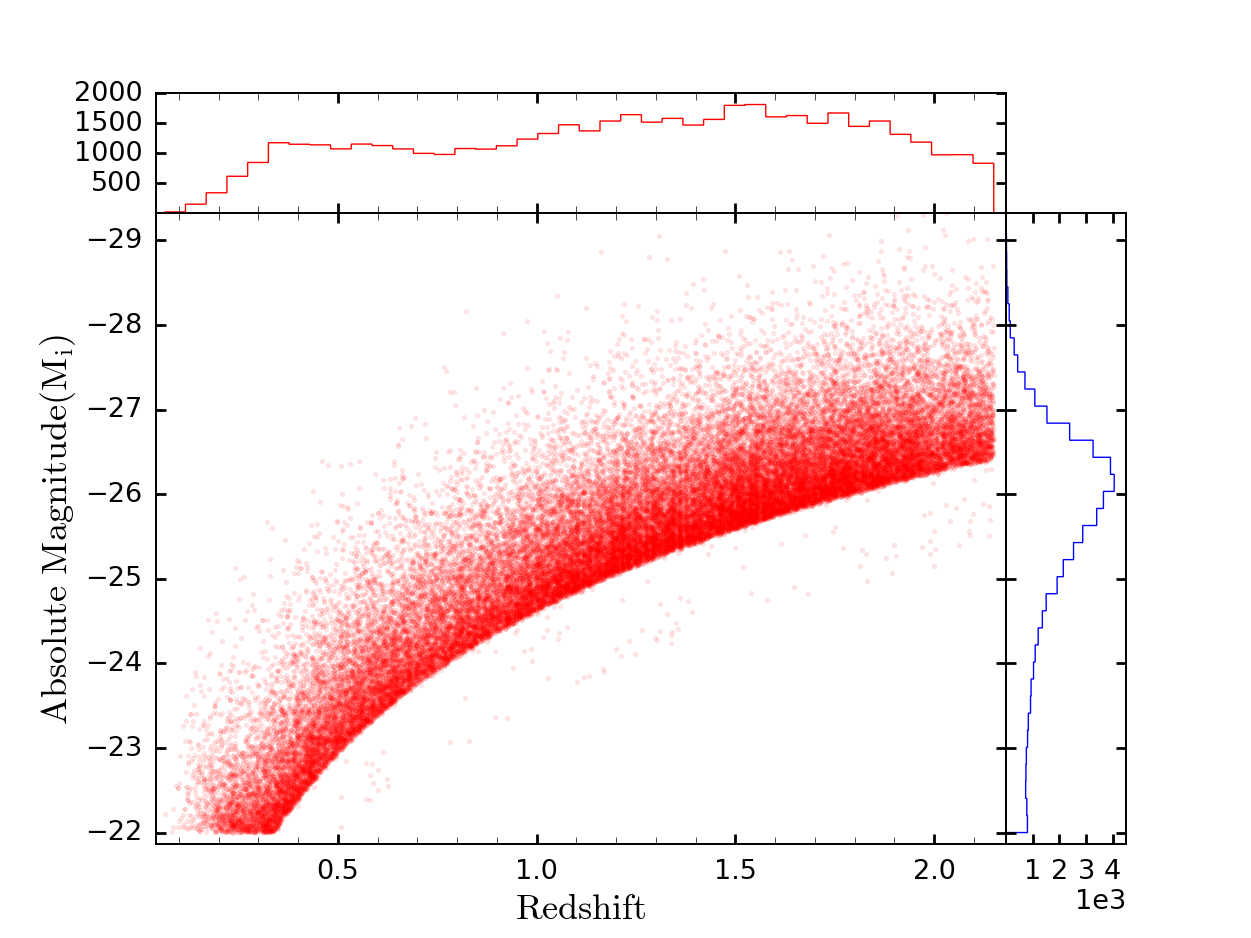}
\caption{The redshift--absolute-magnitude distribution of the uniformly-selected SDSS quasars (\protect\citealt{Richards-2002}). The upper and right panels are the histograms of the redshift and K-corrected absolute $i$-band magnitude respectively with bin sizes of $\Delta z = 0.05$ and $\Delta {\rm mag} = 0.2$.}
\label{fig:red}
\end{figure}

\section{Data} \label{sec:data}
In a stacking experiment, where we try to extract information from undetected sources in a given survey, one needs data from another survey in which the sources have already been identified. In this paper we use optically-selected quasars from the SDSS and radio data from the FIRST survey.
 
\subsection{The optical quasar sample}
The optical data are drawn from the quasar catalogue \citep{Schneider-2010} of the SDSS seventh data release (DR7, \cite{Abazajian-2009}). In SDSS, quasars are mainly identified using colour selection for objects in the magnitude range $15 < i < 19.1$ \citep{Richards-2002,Richards-2006}. Quasars are then differentiated from galaxies and stars by their unique colours in multi-dimensional colour-colour space \citep{Fan-1999}: SDSS's candidate quasars are primarily outliers from stellar regions in colour-colour space \citep{Richards-2001b}, and the regions having large stellar contamination were avoided. The quasar sample includes additional sources that are selected because they have a FIRST counterpart. A source is targeted for spectroscopic follow-up if  it is within 2\,arcsec from a source in the FIRST catalogue. The final catalogue contains 105,783 spectroscopically-confirmed quasars, all brighter than $M_i = -22$ with at least one emission line with full width at half-maximum greater than 1,000\,km/s or a relevant absorption feature. 

We use a subsample consisting of 59,932 quasars selected across the survey area (purely colour-selected sources with the flag \textsc{UNIFORM=1} from \citealt{Shen-2011}) for the purpose of having a homogeneous sample of quasars. This sample covers an effective area of 6,248 deg$^2$ \citep{Shen-2011}. \F{fig:red} shows the distribution of sources in absolute magnitude and redshift.

We divide the sources into seven redshift bins (see Table~\ref{table:slices}) reducing the total to 48,046 sources. Since each redshift bin has a non-negligible width, we apply an absolute-magnitude cut to each redshift bin (corresponding to a minimum luminosity cut per bin) to ensure all quasars in the bin are observed within the sensitivity limit of the survey. This reduces the total number of sources to 24,003. The maximum absolute magnitude in each redshift bin corresponds to the optical flux limit at the highest redshift in that bin given by 
\begin{equation}
M_i = m_{i} - 5\log_{10}[\textrm{d}_L(z_{\rm{up}})/10] - K(z),
\end{equation}
where $m_i = 18.7$ is just above the magnitude completeness limit ($m_i = 19.1$) for DR7, $\rm{d}_L(z_{\rm{up}})$ is the luminosity distance (in pc) at the upper redshift of the bin and $K(z)$ is the  $K$-correction from \cite{Richards-2006}.

\begin{figure}
\centering
\includegraphics[width=0.5\textwidth]{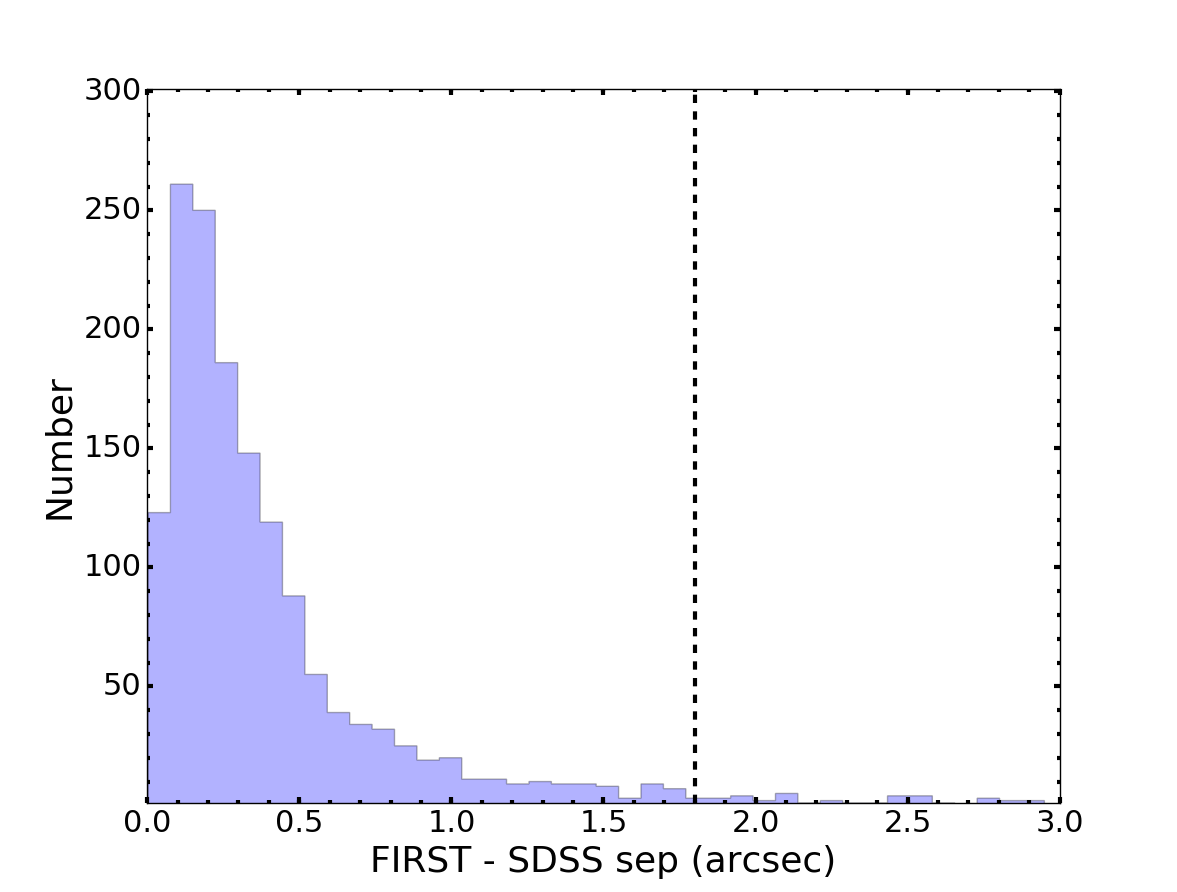}\\ 
\caption{The distribution of the separation between detected FIRST and SDSS quasar positions with a bin size of 0.07\,arcsec. The vertical dashed line at 1.8\,arcsec is the cut-off separation between FIRST and SDSS detected sources that we used in this work.}
\label{fig:astro}
\end{figure}

\subsection{Radio data}
The FIRST survey (\citealt{Becker-1995}) was carried out with the Very Large Array (VLA; \citealt{Thompson-1980}) in its `B' configuration at 20~cm (1.4~GHz), yielding a synthesized beam size of $5.4''$ (FWHM). It covered 8,444 deg$^2$ in the North Galactic cap and 2,131 deg$^2$ in the South Galactic cap giving a total coverage of 10,575 deg$^2$. The survey footprint overlaps with the area that SDSS covered in the North Galactic Cap, as well as with a smaller $\simeq$ 2.5 deg$^2$ wide strip along the Celestial equator. The maps have a rms of $\approx 150\,\mu$Jy/beam. The survey catalogue contains more than 800,000 sources above the detection limit of 1 mJy, and includes peak and integrated flux-densities calculated by fitting a two-dimensional Gaussian to each source. The survey is 95 per cent complete at 2 mJy and 80 per cent complete at 1 mJy. The maps are stored as FITS images and have 1.8$''$ pixels.

\subsection{Cross-matching catalogues}
\label{sec:cross-matching}

We first matched the SDSS quasars with detected sources from the FIRST catalogue. The allowed separation between the coordinates of the two catalogues should be as small as possible to avoid random matching with other sources, but also large enough to ensure real matches are not omitted because of slight random offsets in position between the optical and radio data. 

Fig.~\ref{fig:astro} shows the results of matching our sample to the FIRST catalogue. 
We choose a limiting separation of 1.8\,arcsec based on Fig.~\ref{fig:astro}, which is the pixel size of the FIRST images. From the original 105,783 quasars we made 3,815 matches ($\sim$3 per cent), which is consistent with the low number of optical-to-radio matches found by \citet{Paris-2012} and \citet{Pairs-2017}. We find 2,381 ($\sim$ 10 per cent) matches from our sample of 24,003 SDSS quasars.

\begin{figure}
\centering
\subfloat{\includegraphics[width=0.5\textwidth]{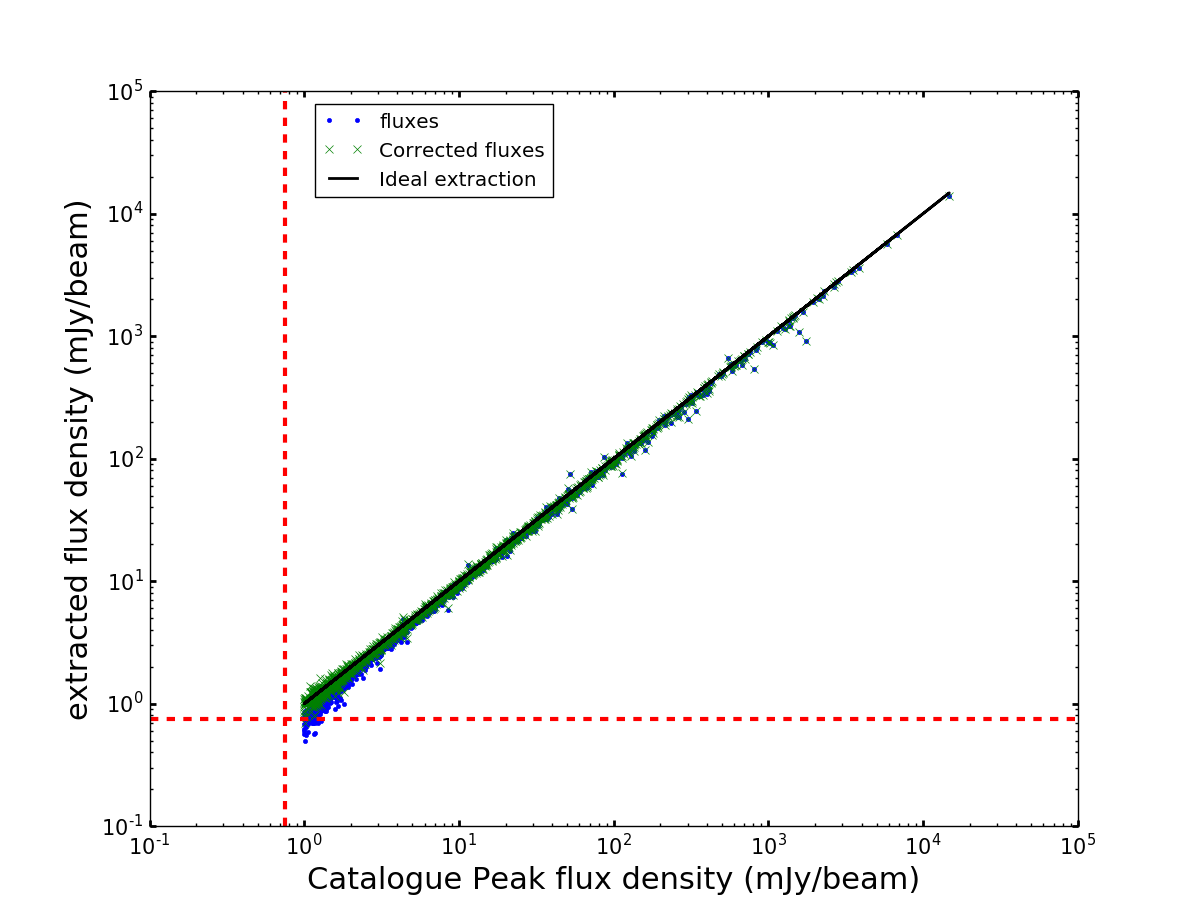}}\\
\caption{Comparison between the FIRST-catalogue peak flux-densities and map-extracted flux-densities, represented by the blue points. The green crosses denote the extracted flux-densities after correction for the biases described by Eq~\ref{eqn:frac}. The solid black line represents the case where the extracted flux-densities would be equal to the catalogue flux-densities, and the dashed red lines indicate the 5$\sigma_n$ threshold.}
\label{fig:xf}
\end{figure}

The FIRST catalogue only contains sources with flux densities above the detection threshold of $\sim 1$\,mJy. In order to obtain sources with flux densities below the FIRST detection threshold we extracted $11\times11$ pixel stamps ($19.8 \times 19.8$\,arcsec) from the FIRST maps, centered on the SDSS quasar positions, and used the central pixel value as the radio flux-density of the quasar. 23,490 of our quasars have fluxes densities, the rest fall outside FIRST coverage.

In Fig~\ref{fig:xf} we compare the catalogued peak flux-densities and the extracted flux-densities for 2,381 detected sources. Most of the extracted flux-densities are in good agreement with peak flux-densities, with the exception of fluxes densities below 10~mJy, which underestimate the peak flux densities. There are also about 10 sources with high scatter from the peak flux-densities. The difference between the extracted flux-densities and peak flux-densities at low flux densities will affect the results and therefore needs to be accounted for and understood. Note, however, that there could be a difference in the effect on extraction of high signal-to-noise (detected) sources compared to the undetected ones. For instance, detected sources could be more extended and therefore slightly resolved by the FIRST restoring beam. Other possible contributions to the difference in the flux densities are \textsc{clean} and snapshot biases. 

\textsc{clean} bias is a systematic effect that decreases the peak flux-density of a source above the detection limit and redistributes it around the map. This phenomenon is associated with the non-linear \textsc{clean} process (\citealt{Condon-1994}) and affects large-area radio surveys such as FIRST and the NRAO VLA Sky Survey (NVSS; \citealt{Condon-1998}). The bias is additive and has an approximately constant magnitude, with a value of 0.25\,mJy\,beam$^{-1}$ for FIRST \citep{Becker-1995}. \cite{White-2007} discovered another bias that affects sub-threshold sources (which are not \textsc{clean}ed) and suggested that it is associated with the sidelobes of the beam pattern. This snapshot bias behaves differently from the one associated with  \textsc{clean} as it is multiplicative (i.e.~the higher the flux density the higher the bias). The proposed total bias correction summarized by \cite{White-2007} is
\begin{equation}
\label{eqn:frac}
S = \min(1.40~S_{\rm F}, S_{\rm F} + 0.25 \rm{\,mJy}),
\end{equation}
where $S$ should be the intrinsic flux-density of the source and $S_{F}$ is the noiseless flux-density from FIRST. This is an idealised case, where $S_{\rm F}$ is meant to incorporate the calibration effects in the FIRST data. It is important to note that this correction can only be applied when the noise can be neglected, that is, for the case of detected sources or the stacked median flux as done in \cite{White-2007}. With the correction, the low (detected) flux-densities are in good agreement with the catalogue flux-densities (Fig~\ref{fig:xf}). Since, in our case, we are also dealing with noise dominated sources, these corrections need to be incorporated directly into the likelihood function in a forward model way, as described in Section~\ref{sec:likelihood}. Throughout the paper, we use $S_m$ to indicate the measured fluxes from FIRST, which will include noise, as opposed to the ideal noiseless case above, that is:
\begin{equation}
\label{eqn:noise}
S_m = S_{\rm F} + n,
\end{equation}
where $n$ represents the noise distribution.

To proceed with the analysis, the sources in each redshift bin in Table~\ref{table:slices}, were further binned in terms of the measured radio flux-density ($S_m$, \F{fig:fluxes}). This includes both detected and undetected sources. One can see from the negative side of the flux-density distributions that the noise is Gaussian to a good approximation, whilst there is a tail on the positive side of the distributions, which shows the contribution from faint real sources. There is an offset in the noisy flux-densities because the average `true' flux-density of these faint sources is comparable to the noise. The flux-density distribution is more Gaussian-like if the noise is much larger than the true flux-density of the faint sources. This effect is more-pronounced for the higher-redshift bins, where a greater fraction of sources are undetected.

\begin{figure*}
\centering
\includegraphics[width=1\textwidth]{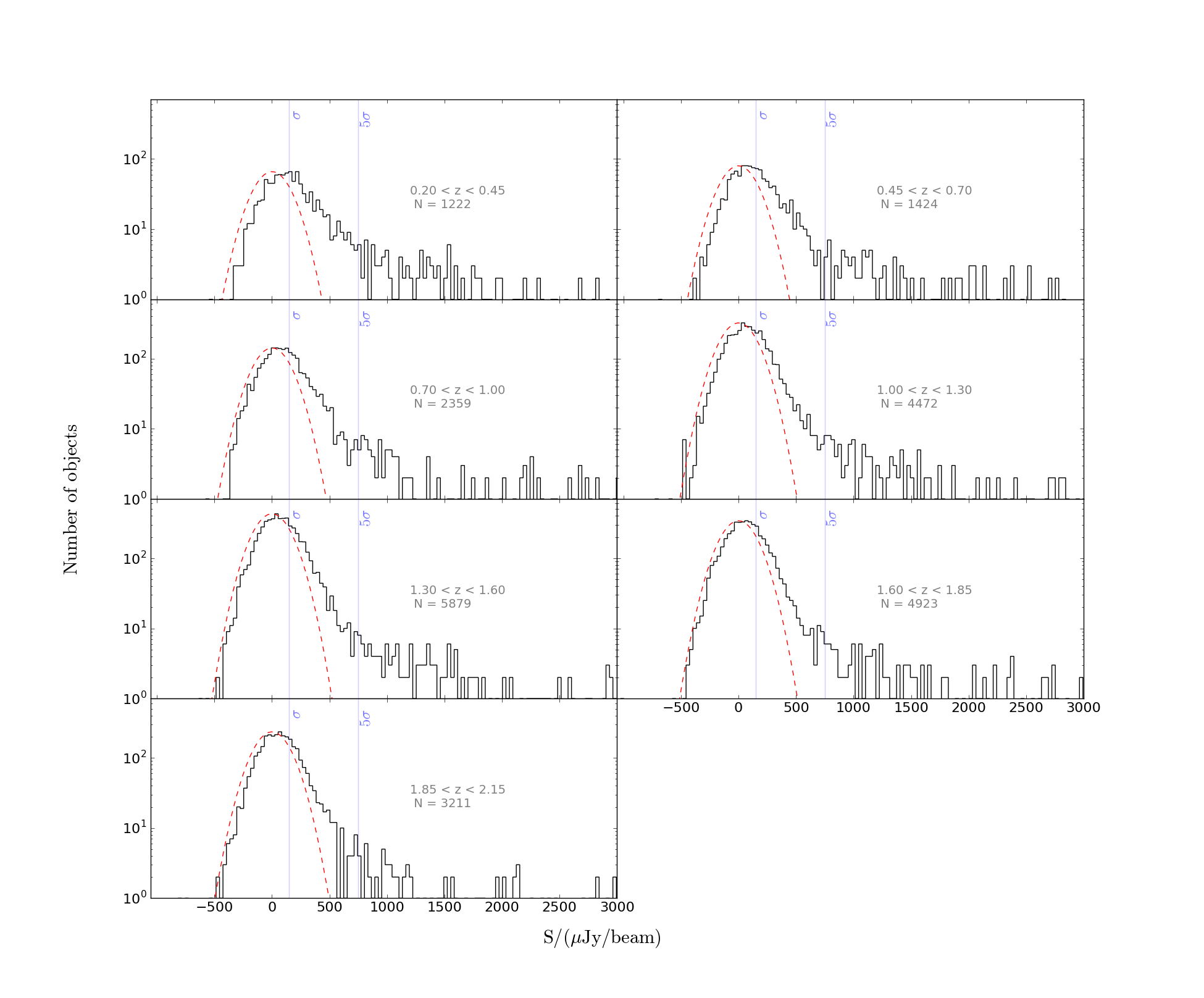}
\caption{Histograms of the raw FIRST flux-densities ($S_m$) extracted from cut-outs centered at the SDSS quasar positions, with $30 \mu$Jy bins. The quasars are divided into 7 redshift bins from Legacy \citep{Shen-2011}. The two blue lines in each bin represent the FIRST rms $\sigma_n = 150 \mu$Jy and $5\sigma_n = 750 \mu$Jy. The red dashed curve is a fixed Gaussian distribution, of mean flux density of zero and $\sigma_n = 150 \mu$Jy, which represents the expected flux-density distribution if there were no sources in the map.}
\label{fig:fluxes}
\end{figure*}
 
\begin{table} 
\centering 
\caption{The redshift bins used to separate the sources, along with the absolute-magnitude cut, the number of sources ($N$), and the number of sources with extracted FIRST flux-densities, $N_{\mathrm{FIRST}}$ (most of these sources are undetected in FIRST). The RQF (`radio-quiet' fraction) of the quasars is calculated by integrating over the low-luminosity part of the radio luminosity function described in Sec~\ref{sec:models}.} 
\label{table:slices} 
\begin{tabular}{ccccc} \hline 
   Redshift bin   &   max($M_i$)&  $N$  &$N_{\mathrm{FIRST}}$ &      RQF (\%) \\ \hline 
$0.20 < z < 0.45$ &   -23.0     & 1234  &   1222     &      96.2 \\ 
$0.45 < z < 0.70$ &   -24.1     & 1437  &   1424     &      91.8  \\ 
$0.70 < z < 1.00$ &   -24.9     & 2401  &   2359     &      86.3 \\ 
$1.00 < z < 1.30$ &   -25.4     & 4534  &   4472     &      93.8 \\ 
$1.30 < z < 1.60$ &   -25.8     & 5967  &   5879     &      93.7 \\ 
$1.60 < z < 1.85$ &   -26.2     & 4988  &   4923     &      91.4 \\ 
$1.85 < z < 2.15$ &   -26.6     & 3250  &   3211     &      91.4\\  
\hline \hline 
\end{tabular} 

\end{table} 


\section{Bayestack framework} \label{sec:Below}

Our stacking analysis is based on a Bayesian formalism that can probe the quasar RLF below the FIRST detection threshold, down to sub-mJy levels. We made use of a modified version of the software \textsc{bayestack} 
\citep{Zwart-2015}. The idea is to start with a model for the RLF for a given redshift bin. We then translate that into a source-count model, take into account the FIRST bias correction, and fit to the number of sources per flux-density bin, as extracted from the data. Below we review the basics of the method.

\subsection{Bayesian analysis}
The fitting approach centres on Bayes' theorem,
\begin{equation}
\centering
\mathcal{P}(\Theta|D,H) =\frac{\mathcal{L}(D|\Theta,H) \Pi(\Theta|H)}{\mathcal{Z}},
\label{eqn:bayes}
\end{equation}
where $\mathcal{P}$ is the posterior distribution of the parameters $\Theta$, given
the data $D$ and model $H$. $\mathcal{L}$ is the likelihood, the probability
distribution of the data given the model and parameters, and $\Pi$ is
the prior, the known constraints on the parameters. $\mathcal{Z}$ is the Bayesian evidence, which normalizes $\mathcal{P}$ and can be written as an integral of $\mathcal{L}$ and $\Pi$ over the $n$-dimensional parameter space $\Theta$,
\begin{equation} 
\mathcal{Z}  = \int \mathcal{L} \Pi \textrm{d}^n\Theta. 
\label{eqn:Z} 
\end{equation} 
A model has high evidence when a large portion of its prior parameter space is likely (i.e.~large likelihood), and small evidence when a large portion of its parameter space has a small likelihood, irrespective of how peaked the likelihood function is. This therefore automatically encapsulates Occam's razor (e.g.~\citealt{Feroz_Hobson-2008}).

In order to compute this posterior distribution, one needs to sample from it. Sampling has always been one of the most computationally expensive parts of model selection because it involves solving the multidimensional integral in Eq.~\ref{eqn:Z}. Nested sampling (\citealt{Skilling-2004})
was created for its efficiency in calculating the evidence, with an added bonus of producing posterior inferences as a by-product.
\nest\, \citep{Feroz_Hobson-2008,Feroz-2009a,Buchner-2014} is a robust implementation of nested sampling, returning the full posterior distribution from which the uncertainty analysis can be correctly undertaken.

In Bayesian model selection, one compares the evidences of two models, A and B. This is quantified by considering the ratio of their evidences $\mathcal{Z}_A/\mathcal{Z}_B$ (equivalent to the difference of their log-evidence, $\ln [\mathcal{Z}_B -\mathcal{Z}_A$]), known as the Bayes factor.  \cite{Jeffreys-1961} introduced a way to conclude how much better Model A is compared to B using the Bayes factor: $\Delta \ln \mathcal{Z} < 1$ is `not significant', $1 < \Delta \ln \mathcal{Z} < 2.5$ is `significant', $2.5 < \Delta \ln \mathcal{Z} < 5$ is `strong', and $\Delta \ln \mathcal{Z} > 5$ is `decisive'. We adopt this scale in our analysis and use it to compare different luminosity models.

\subsection{Assumed Likelihood \label{sec:likelihood}}

To proceed with our Bayesian analysis we need a likelihood for the data, which in our case comprises the extracted flux-densities, $S_m$. As explained in Section~\ref{sec:cross-matching}, this flux density is a combination of the biased FIRST flux density ($S_{\rm F}$) of the source and the noise distribution (\Eq{eqn:noise}). The noise is assumed to follow a Gaussian distribution, centered at zero with a variance $\sigma_n^2$. This is a good assumption considering that the flux-density distribution of source-less (`random sky') FIRST extracted flux-densities is well approximated by a Gaussian. 

Since we are working with binned flux-densities, the likelihood of finding $k_{i}$ objects in the $i^{th}$ flux-density bin [$S_{mi}, S_{mi} + \triangle S_m$] follows a Poisson distribution,
\begin{equation} 
\mathcal{L}_{i}\left(k_i|\pmb{\Theta}\right) = \frac{I_i^{k_i}\me^{-I_i}}{k_i!}, 
\end{equation} 
where $I_i$ is the theoretically-expected number of sources in the $i^{th}$ measured bin, $[S_{m_i} ,S_{m_i}+ \Delta S_{m} ]$, given by the modified equation taken from \cite{Mitchell_Wynne-2013},
%
\begin{equation} 
\label{eqn:ii} 
I_i= 
\int_{S_{min}}^{S_{max}} 
dS \frac{dN(S)}{dS} 
\int_{S_{m_i}}^{S_{m_i}+\Delta S_{m_i}} 
dS_m 
\frac{1}{\sigma_{n}\sqrt{2\pi}} 
\rm{e}^{-\frac{\left(S_{\rm F}-S_m\right)^2}{2\sigma_n^2}}.
\end{equation} 
Here $dN/dS$ is the source-count model (number of sources per flux-density bin), $\sigma_n$ is the mean noise of the data and $S$ is again the intrinsic flux density of the source. The noiseless FIRST flux density $S_{\rm F}$ is related to $S$ by \Eq{eqn:frac}. Therefore, in order to compare to the measured flux, we need to first apply the expected bias effects of the FIRST observation, e.g. $S_{\rm F}=\max\{S/1.4, (S - 0.25 \rm{\,mJy})\}$. This approach naturally takes into account sample variance (at the Poisson level) since it does not fix the total number of predicted sources to the observed number (e.g. other regions of the sky could have a different total number). This will have implications for the allowed minimum and maximum flux-density values of our fits, as we will see later. The fitting will have large variance at the low flux-density level (because of the noise) and at the high flux-density level (because of Poisson fluctuations due to the low number of sources).
Solving the second integral, \Eq{eqn:ii} becomes 
%
\begin{equation} 
\label{eqn:iii} 
\begin{aligned}
I_i= & \int_{S_{min}}^{S_{max}} dS \frac{dN(S)}{dS}\\ 
     &\frac{1}{2} \left\{\mathrm{erf}\left(\frac{S_{\rm F}-S_{m_i}}{\sigma_n\sqrt{2}}\right) 
- \mathrm{erf}\left(\frac{S_{\rm F}-(S_{m_i}+\Delta S_{m_i})}{\sigma_n\sqrt{2}}\right)\right\}.
\end{aligned}
\end{equation} 

The total likelihood for the $N$ bins is given by the product of the likelihood in each bin, assuming that the bins are independent, 
\begin{equation} 
\label{eqn:lhood-tot-bins} 
\mathcal{L}\left(\mathbf{k}|\pmb{\Theta}\right) 
    =\prod_{i=1}^{N} \mathcal{L}_{i}\left(k_i|\pmb{\theta}\right). 
\end{equation}
As we aim to fit models that describe the radio luminosity function, we need to convert those luminosity-function models to source counts, $dN/dS$, and compare to the binned flux-densities in data space where the noise is Gaussian. As a final detail, we would like to point out that we include bins with zero sources at the low flux-density (negative) end. This means we do not actually see any galaxies below a certain flux-density level (including noise) and models that predict galaxies in those flux-density bins should be penalized. At the high flux-density level this is not done as the maximum flux-density cutoff is our choice, and models that predict some sources above that should not be penalized. However, such models will likely over-predict sources in our highest flux-density bin and will therefore have a lower probability. In any case, such a choice has very little impact on the low-flux-density stacked sample that we are targeting in our analysis.

\subsection{Models for the radio luminosity functions \label{sec:models}}

The luminosity per unit frequency (luminosity density) of a radio source, $L_\nu$, can be related to the observed flux-density at the same frequency, $S_\nu$, through 
\begin{equation}
L_\nu = 4 \pi D_L^2( 1+z)^{\alpha-1}S_\nu,
\end{equation}
where $D_L$ is the luminosity distance, $\alpha$ is the spectral index of the source, and $z$ is the redshift of the source.

The luminosity function (LF), $\rho(L_\nu)$, is the number density of sources per luminosity density bin, e.g. $\rho(L_\nu)=dN/(dL dV)$ (where $dV$ is comoving volume).
Another common definition of the LF ($\Phi$), which we use here, involves binning the source counts in magnitude, that is, $m-m_0=-2.5 \log_{10}(L/L_0)$. The relationship between these two definitions is then
\begin{equation}
\Phi (L_\nu) = \frac{dN}{dV dm} = \frac{dN}{dV dL_\nu}\frac{dL_\nu}{dm} = \ln(10^{0.4}) L_\nu \rho (L_\nu).
\end{equation}
We define parametric models for the quasar RLF consisting of two functions, one for the luminous sources and the other for faint sources (using subscripts 1 and 2 respectively). The radio-loud quasar RLF has been shown to follow a double power-law (see e.g.~\citealt{Boyle-1988}), so we parameterize the luminous part of the RLF as a double power-law for all the models considered here. The shape of the quasar RLF at low luminosities is still uncertain, so for that we consider 3 models: a power-law, a double-power-law and a log-normal power-law. 

\textbf{Model A} is the simplest overall form for the quasar RLF -- a double power-law for the high luminosities (the detected sources) and a single power-law to describe the RLF at low luminosities:
\begin{equation}
\Phi(L)_A = \frac{\Phi_1^*}{(L/L_1^*)^{\alpha_1} + (L/L_1^*)^{\beta_1}} + \frac{\Phi_2^*}{(L/L_2^*)^{\alpha_2} }.
\label{eqn:dpl_pl}
\end{equation}
Note that $L_2^*$ and $\Phi_2^*$ will be degenerate here, but we keep this form for convenience. 

\textbf{Model B} has a double power-law for both the high- and low-luminosity sources:
\begin{equation}
\Phi(L)_B = \frac{\Phi_1^*}{(L/L_1^*)^{\alpha_1} + (L/L_1^*)^{\beta_1}} + \frac{\Phi_2^*}{(L/L_2^*)^{\alpha_2} + (L/L_2^*)^{\beta_2}}.
\label{eqn:dpl_dpl}
\end{equation}

\textbf{Model C} has a double power-law for the luminous sources and a log-normal power-law, which has earlier been used for star-forming galaxies (\citealt{Tammann-1979}), for low-luminosity sources:
\begin{equation}
\begin{aligned}
\Phi(L)_C =& \frac{\Phi_1^*}{(L/L_1^*)^{\alpha_1} + (L/L_1*)^{\beta_1}} \\
        &+\Phi_2^* \left( \frac{L}{L_2^*} \right)^{1-{\delta}} \exp \left[-\frac{1}{2\sigma_{LF}^2} \log_{10}^2 \left(1 + \frac{L}{L_2^*}\right) \right].
\end{aligned}
\label{eqn:dpl_lognorm}
\end{equation}

Finally, we note that each of the model functions will be bounded: $L_{{\rm min}_1} \le L \le L_{{\rm max}_1}$ for the high-luminosity end and $L_{{\rm min}_2} \le L \le L_{{\rm max}_2}$ for the low-luminosity end. The boundaries are allowed to overlap since there might be a contribution from both populations. We also consider a different set of models when fitting to simulations which include only the part used for the low luminosity region (parameters with subscript "2"). We call it \textbf{Model $A'$} (single power law), \textbf{Model $B'$} (one double power-law) and \textbf{Model $C'$} (a log-normal power-law).

The likelihood (\Eq{eqn:iii}) is computed in flux-density space, which means that our LF models, $\Phi(L)$, have to be converted into source-count models, $dN/dS$:
\begin{equation}
\begin{aligned}
\frac{dN}{dS}   &= \frac{dN}{dL} \frac{dL}{dS}\\
                &= \rho(L)  V_i 4\pi D_L^2(1+z_i)^{\alpha-1}\\
                &=  \frac{\Phi(L) V_i}{L\ln(10^{0.4})}
                4\pi D_L^2(1+z_i)^{\alpha-1},\\
\end{aligned}
\end{equation}
where $V_i$ is the volume of the survey for the redshift bin $i$ and $z_i$ is the mean redshift for that bin.

\subsection{Priors}
Priors play an important role in Bayesian inference as they define the sampled parameter space. A uniform prior is the simplest form, providing an equal weighting of the parameter space. We assign a uniform prior to the slopes $\alpha_{1,2}$, $\beta_{1,2}$, and $\delta$. $\sigma_{LF}$ also has a uniform prior. To avoid degeneracy in the slopes for the double power law, we also impose $\alpha_{1,2} \ge \beta_{1,2}$. $L_{1,2}^*$, $L_{{\rm min}_{1,2}}$, $L_{{\rm max}_{1,2}}$ and $\phi_{1,2}^*$ all have uniform priors in log-space. 
The priors are summarised in Table ~\ref{table:prior}.
 
Combining \Eq{eqn:lhood-tot-bins} with the priors shown in Table~\ref{table:prior}, and substituting into \Eq{eqn:bayes}, one can determine the posterior probability distribution as well as the evidence. We use a Python implementation (\citealt{Buchner-2014}) of \nest\, (Py\nest) to fit the models with evidence\_tolerence=0.5 and sampling\_efficiency=0.1.  

\begin{table} 
\centering 
\caption{Assumed priors. $L_{5\sigma}$ is the luminosity corresponding to the $5\sigma_n$ flux-density cut for a given redshift.}
\label{table:prior} 
\begin{tabular}{ll} 
\hline 
Parameter  & Prior   \\ 
\hline 
$\alpha_1,\beta_1,\alpha_2, \beta_2,\delta$           & uniform $\in \left[-5,5\right]$  \\ 
$\sigma_{LF}$                                         & uniform $\in \left[0.1,2\right]$  \\ 
$\log_{10}[L_{\rm{min_{\{1,2\}}}}/(\rm{WHz}^{-1})]$          & uniform $\in \left[20,30\right]$ \\ 
$\log_{10}[L_{\rm{max_{\{1,2\}}}}/(\rm{WHz}^{-1})] $         & uniform $\in \left[20,30\right]$ \\ 
$\log_{10}[\phi_{\{1,2\}}^*/(\rm{Mpc}^{-3} \rm{mag}^{-1})]$ & uniform $\in \left[-12,-2\right]$ \\
$\log_{10}[L_1^*/(\rm{WHz}^{-1})]$                   & uniform $\in \left[\log_{10}(L_{5\sigma}),30\right]$ \\
$\log_{10}[L_2^*/(\rm{WHz}^{-1})]$                    & uniform $\in \left[20,\log_{10}(L_{5\sigma})\right]$ \\ 
\hline 
\end{tabular} 
\end{table} 

\section{Tests on simulated data} \label{sec:Test}

We first test our technique by applying it to the Square Kilometre Array Design Studies SKA Simulated Skies (SKADS-S$^3$) simulations (see \citealt{Wilman-2008,Wilman-2010}). SKADS is a semi-empirical simulation of the extragalactic radio continuum sky, covering a sky area of $20 \times 20$ deg$^2$ with $\approx 320$ million sources out to a redshift of $z = 20$ and flux density of 10\,nJy. 

We took $\sim 555,000$ sources contained within a 8\,deg$^2$ patch of the simulation in the redshift range $1.0 < z < 1.3$. 223,457 of those sources have radio luminosities between $20.5 < \log_{10}[L/{\rm W\,Hz}^{-1}] < 24.5$ and we call this the full sample.
In order to test how a higher luminosity cut may alter our fits, we also consider a brighter sample of 91,458 which lie between $21.5 < \log_{10}[L/{\rm W\,Hz}^{-1}] < 24.5$. Such luminosity cut could arise due to the input optical sample being being flux limited, and if there is a correlation between the optical emission and radio emission, this would in turn lead to a downturn in the measured RLF that may not happen if one could measure it directly from a purely radio-selected sample.

We added random noise -- generated from a Gaussian distribution, with standard deviation $\sigma_n$ that corresponds to the FIRST rms of 150 $\mu$Jy ($S_n = S + N[150$ $\mu$Jy, 0]) -- to both the high luminosity sample ($\log_{10}[L/(\rm{WHz}^{-1})]> 21.4$) and the full sample ($\log_{10}[L_{1.4}/{\rm W\,Hz}^{-1}]>20.4$), to emulate the observed data (i.e.~the `noisy' 
sources in FIRST). We bin the noisy SKADS sources in flux density and apply our technique for fitting the three models (setting $S=S_{\rm F}$ as there are no calibration biases in the simulated data). In this case we only fit a single function from each model (either a power-law, double power-law or log-normal) to the faint SKADS sources, to test the technique on sources around and below the detection threshold (what we call model $A'$, $B'$ and $C'$). We repeat this using the extreme case of a noise level of 15 $\mu$Jy which will allow for detected sources. 

We note that with real data, if the parent catalogue (in the case of the SDSS quasar sample we use in this paper) is flux limited, this may naturally lead to a lower limit in radio luminosity that we can probe, if there is a correlation between optical and radio luminosity \citep[e.g.][]{Serjeant-1998, White-2017}. 

{\nest} returns the Bayesian evidence of the model and the posterior distribution for all the fitted parameters. The `relative evidence' for a model is the difference between the model evidence and the reference-model evidence (where the reference model is the model with the lowest evidence). We show the relative evidence for the SKADS samples in Table~\ref{table:skads}, where the winning model, the one with the highest relative evidence, is in bold. From the relative evidences it is clear that the data prefers the log-normal function (Model $C'$) for the 15~$\mu$Jy noise levels in both samples and the power-law (Model $A'$) for the 150$~\mu$Jy noise levels, although the evidence is marginal between models for the 150$~\mu$Jy noise level. The evidence also suggests that the power-law function (Model $A'$) is a significantly poor fit compared to the other models for the 15\,$\mu$Jy noise-levels.

In \F{fig:tri_skads} we show the one-dimensional (1-D) and two-dimensional (2-D) posterior distributions for the fits of the various models to the `noisy' low-luminosity SKADS sample. The 1-D posterior distribution is the marginalization for each parameter, located at the end of each row in \F{fig:tri_skads}. The peaks in 1D do not always do justice to the 2-D posteriors as they are not just simple Gaussians. They show distorted `banana-like' shapes, with some having long tails. The limits on each plot are the maximum and minimum values from the posterior distribution. Some of these parameters are unconstrained and therefore limited by the assumed priors. Note, however, that this has no effect on the  reconstructed RLFs.

Along with the posterior distribution, {\nest} returns three values to summarize each parameter: the mean, maximum-likelihood and maximum-a-posteriori (MAP, maximizing the product of the likelihood and prior) values. Obtaining a single value for a parameter is straightforward if the 1-D posterior is Gaussian, as the mean, MAP and maximum likelihood are the same or very close to each other. It is clear that some of
the posteriors in \F{fig:tri_skads} are not Gaussian, which would mean that the three summaries are likely to be different from each other. We use all three parameters to reconstruct the LFs in turn. (The maximum likelihood gives the same value as the MAP for models with uniform priors, so we just quote the MAP.) Although they are good estimates, they still do not fully describe the complex nature of the posterior, as clearly shown in \F{fig:tri_skads}.

In \F{fig:skads} we show the reconstruction of radio luminosity functions (RLFs) of the noisy SKADS sources, using both the $21.5 <\log_{10}[L_{1.4}/{\rm W\,Hz}^{-1}] < 24.5$ and $20.5 <\log_{10}[L_{1.4}/{\rm W\,Hz}^{-1}] < 24.5$ samples. We also show the average total RLF, MAP functions and the 95 per cent confidence interval for each model fit and noise level. Such a choice is not unique as the models that span the 95 per cent confidence interval do not necessarily give a continuous region in terms of the RLF curves. For plotting such a region, we chose a set of luminosity bins and calculated all the values of the RLFs in each bin corresponding to all the models in the posterior to determine the 95 per cent confidence limits.

\begin{figure*}
\centering
\includegraphics[width=0.8\textwidth]{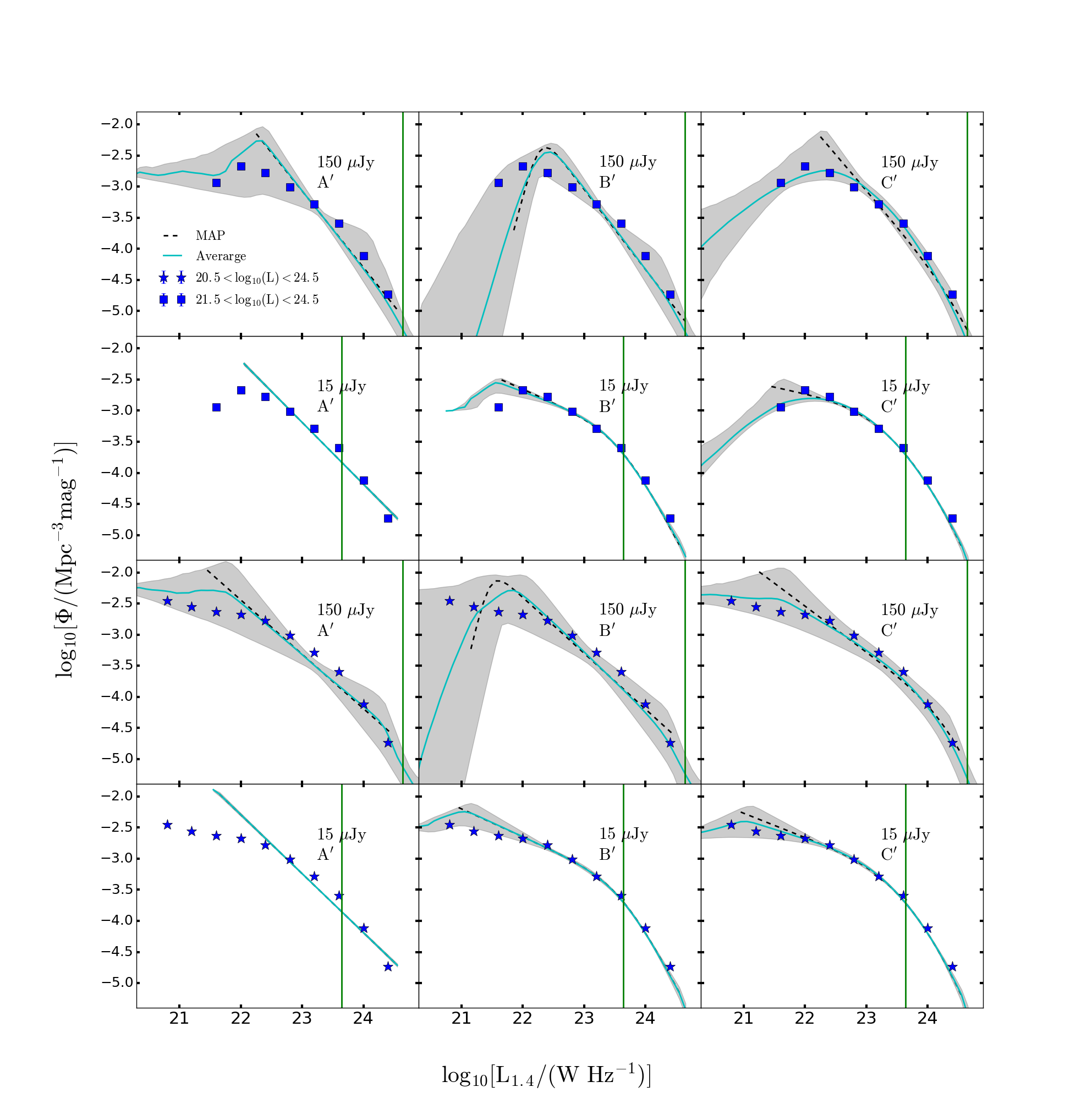}\\
\caption{The SKADS radio luminosity function and the reconstruction of the RLF using \textsc{bayestack} for the $21.5 <\log_{10}[L/(\rm{WHz}^{-1})] < 24.5$ (two top panels) and $20.5 <\log_{10}[L/(\rm{WHz}^{-1})] < 24.5$ (bottom two panels) samples. The top panels and third panels from the top are the reconstructions when noise of 150-$\mu$Jy is added to the simulated SKADS sources. The second panels and bottom panels use 15-$\mu$Jy noise. The left panels are the power-law models ($A'$), the middle panels are the double power-law models ($B'$) and the right panels are the log-normal models ($C'$). The blue squares and blue stars denote the true SKADS RLFs for the $21.5 <\log_{10}[L/(\rm{WHz}^{-1})] < 24.5$ and $20.5 <\log_{10}[L/(\rm{WHz}^{-1})] < 24.5$ samples respectively. The cyan and black dashed curves respectively represent the RLFs reconstructed using the mean and MAP parameters. The grey regions represent the 95-per-cent confidence intervals for the distribution of model reconstructions in the posterior. The green solid line represents the 5$\sigma_n$ noise shown in the top-right corner of each panel.}
\label{fig:skads}
\end{figure*}

Since this is a simulation, we can calculate the true underlying RLF by converting flux density to luminosity and bin in luminosity and volume (given the sky area and redshift bin). We therefore show the comparison to these RLFs in \F{fig:skads}. We see that the reconstructed RLFs with 150-$\mu$Jy noise levels have a large scatter but are in good agreement with the true SKADS RLF. As expected, using lower noise levels produces RLF reconstructions with better fits to the SKADS RLF and smaller 95-per-cent confidence regions (the 15\,$\mu$Jy noise level panels in Fig.~\ref{fig:skads}). Thus the fitting method works for our current noise levels and for those that will be obtained by future radio surveys. We see that the fit is unbiased, though (of course) if the model is quite poor, the fitting will also be poor (as in the case of the power law). Moreover, the fitting is not affected when we use the sample that includes sources down to $\log_{10}[L/(W\,Hz^{-1})] = 20.5$, although (as one would expect) the uncertainty increases as we move to lower luminosities.

\begin{table} 
\centering 
\caption{The relative $\log_{10}-$evidence, $\Delta\log_{10}Z$, for the models using single radio luminosity functions (what we call Model $A'$, $B'$ and $C'$) relative to the model with the lowest evidence in each case, applied to the full SKADS sample and the high-luminosity sample with different noise levels.}
\label{table:skads} 
\begin{tabular}{l|l|l} 
\hline
& 15 $\mu$Jy& 150 $\mu$Jy\\
\hline 
Model & $\Delta\log_{10}Z$ & $\Delta\log_{10}Z$\\ 
\hline
\multicolumn{3}{c}{$21.5 < \log_{10}[L/(\rm{WHz}^{-1})] < 24.5$}\\
\hline
$A'$ & $      0.0 \pm 0.00$  & $\tb{0.45 \pm 0.17}$ \\
$B'$ & $    139.1 \pm 0.22$  & $    0.00 \pm 0.00$  \\ 
$C'$ & $\tb{141.8 \pm 0.21}$ & $    0.35 \pm 0.17$ \\ 
\hline
\multicolumn{3}{c}{$20.5 < \log_{10}[L/(\rm{WHz}^{-1})] < 24.5$}\\
\hline
$A'$ &    $0.00 \pm 0.00$  & $\tb{0.87 \pm 0.14}$ \\
$B'$ &    ${141.5 \pm 0.20}$ & $    0.00 \pm 0.00 $ \\ 
$C'$ & $\tb{147.2 \pm 0.22}$ & $   {0.71 \pm 0.14}$ \\ 
\hline 
\end{tabular} 
\end{table}

\begin{figure*}
\centering
\subfloat[Model $A'$; 150 $\mu$Jy]{\includegraphics[width=0.5\textwidth]{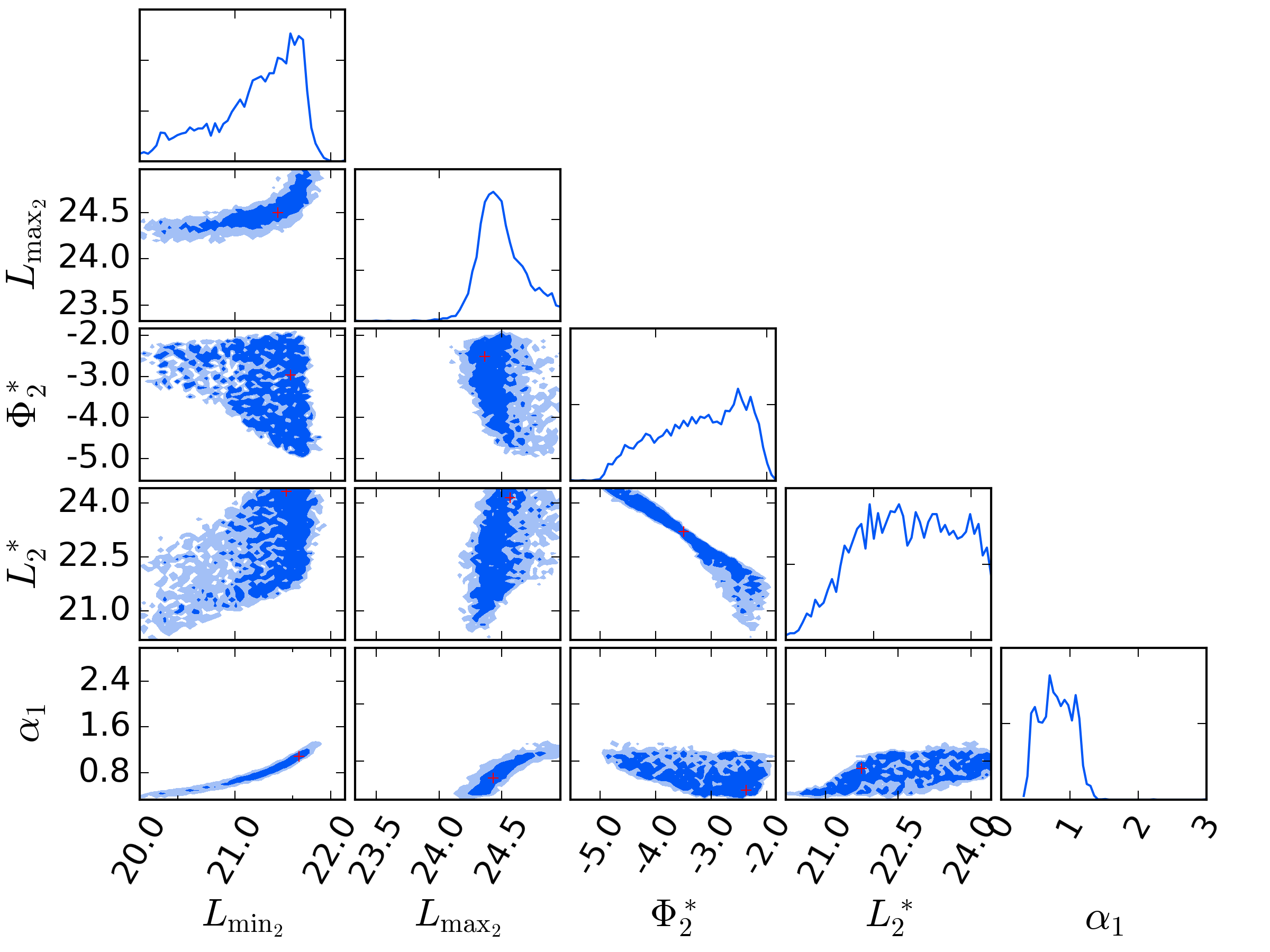}}
\subfloat[Model $A'$;  15 $\mu$Jy]{\includegraphics[width=0.5\textwidth]{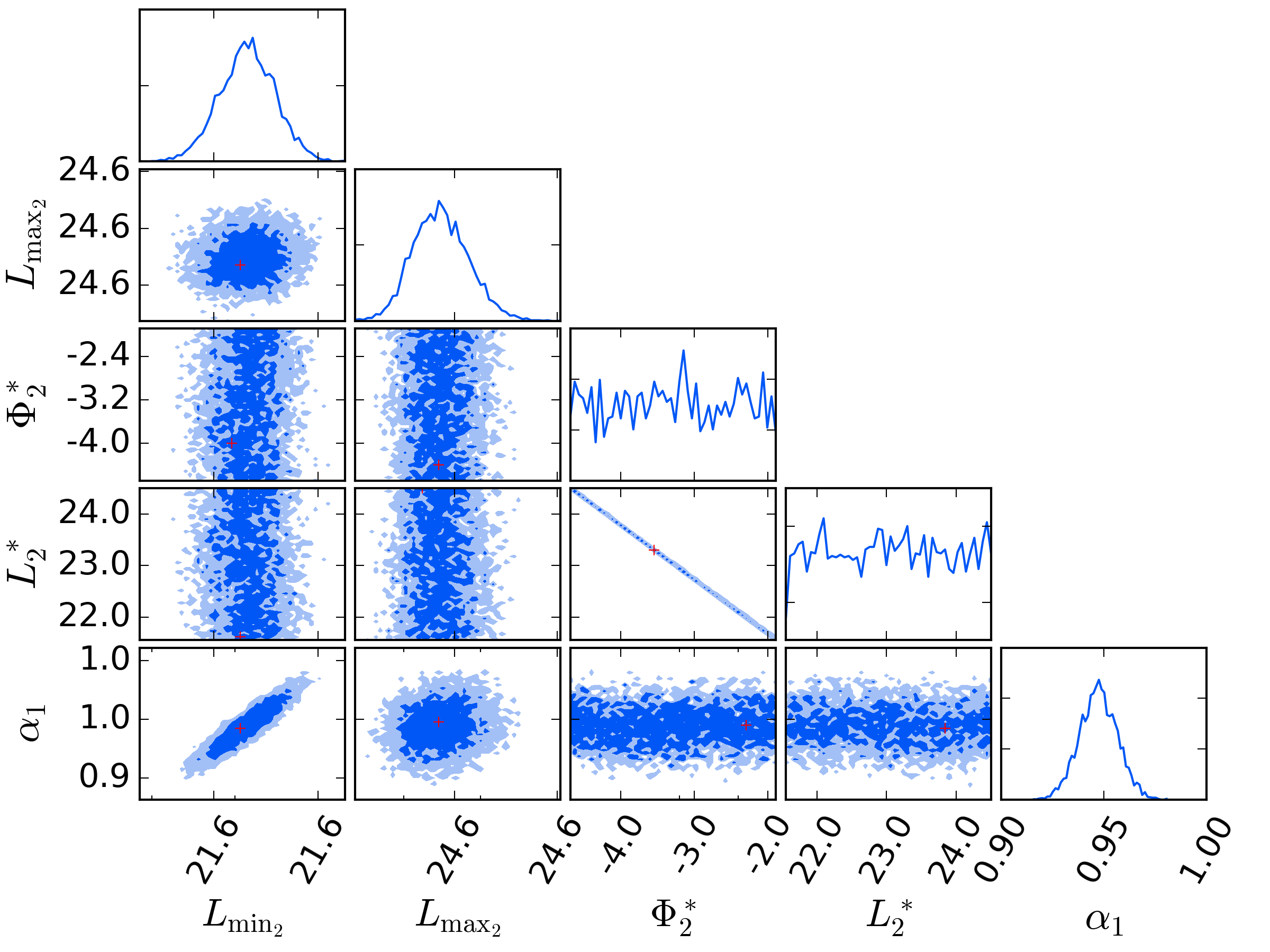}}\\
\subfloat[Model $B'$; 150 $\mu$Jy]{\includegraphics[width=0.5\textwidth]{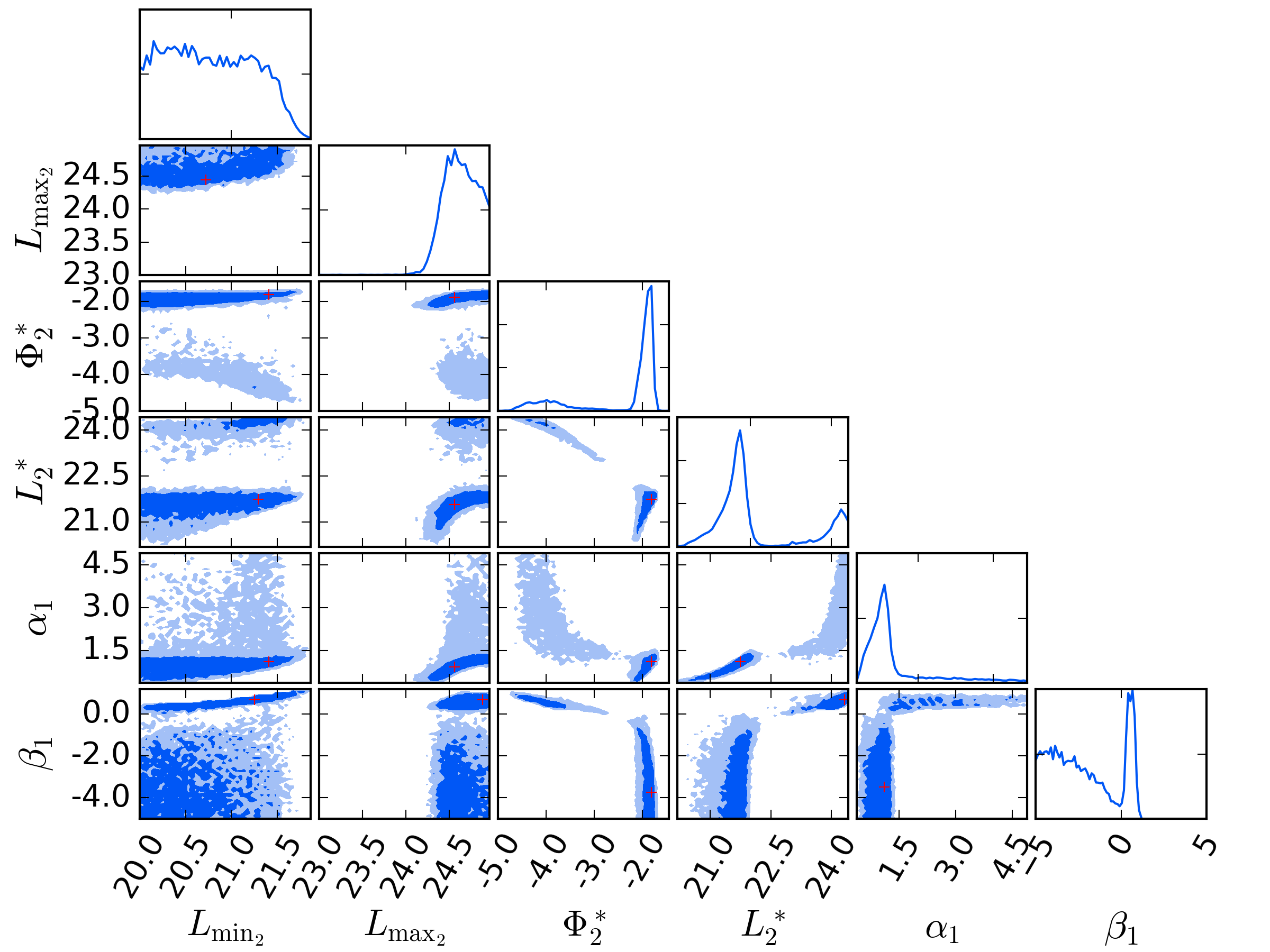}}
\subfloat[Model $B'$;  15 $\mu$Jy]{\includegraphics[width=0.5\textwidth]{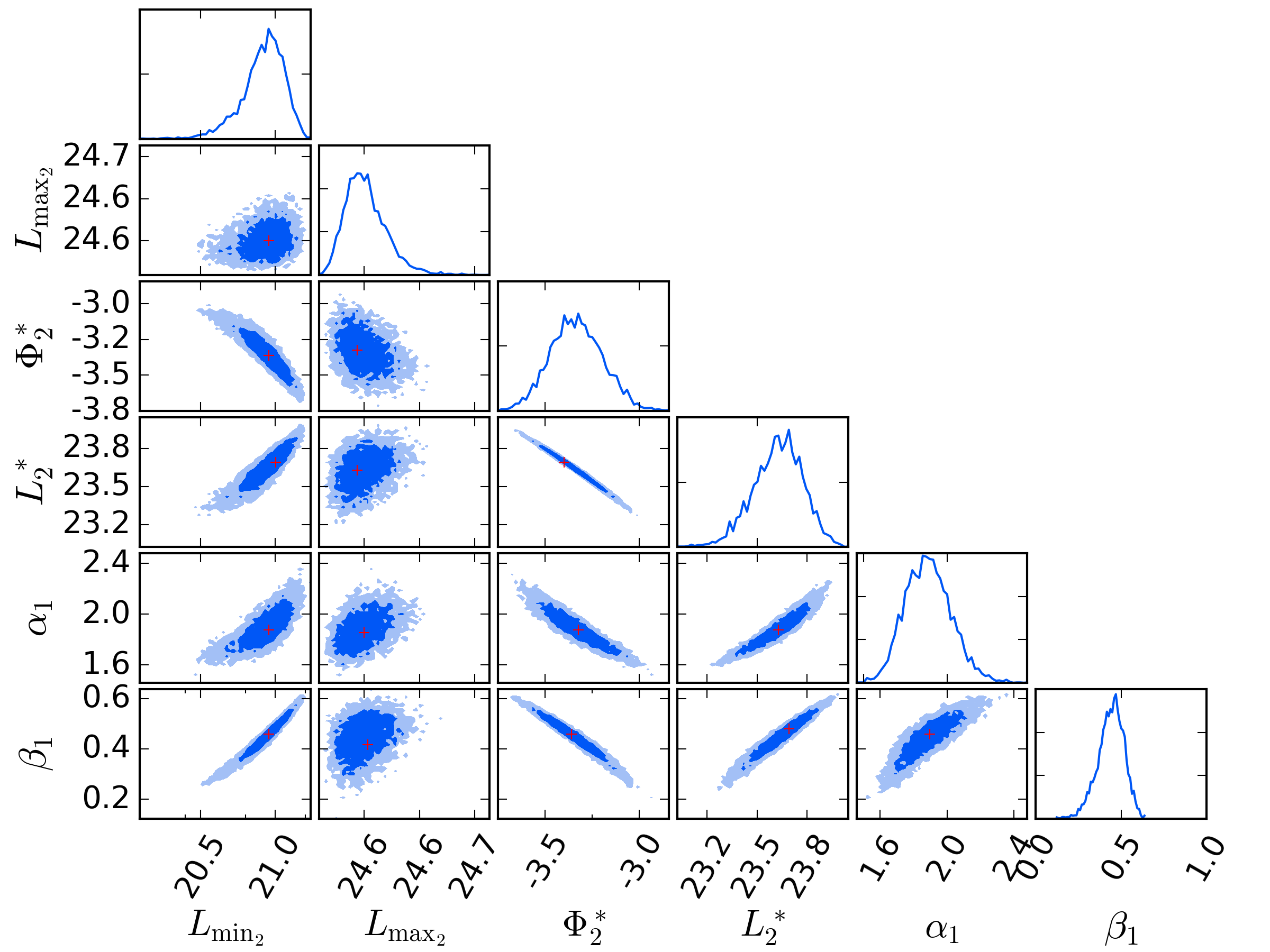}}\\
\subfloat[Model $C'$; 150 $\mu$Jy]{\includegraphics[width=0.5\textwidth]{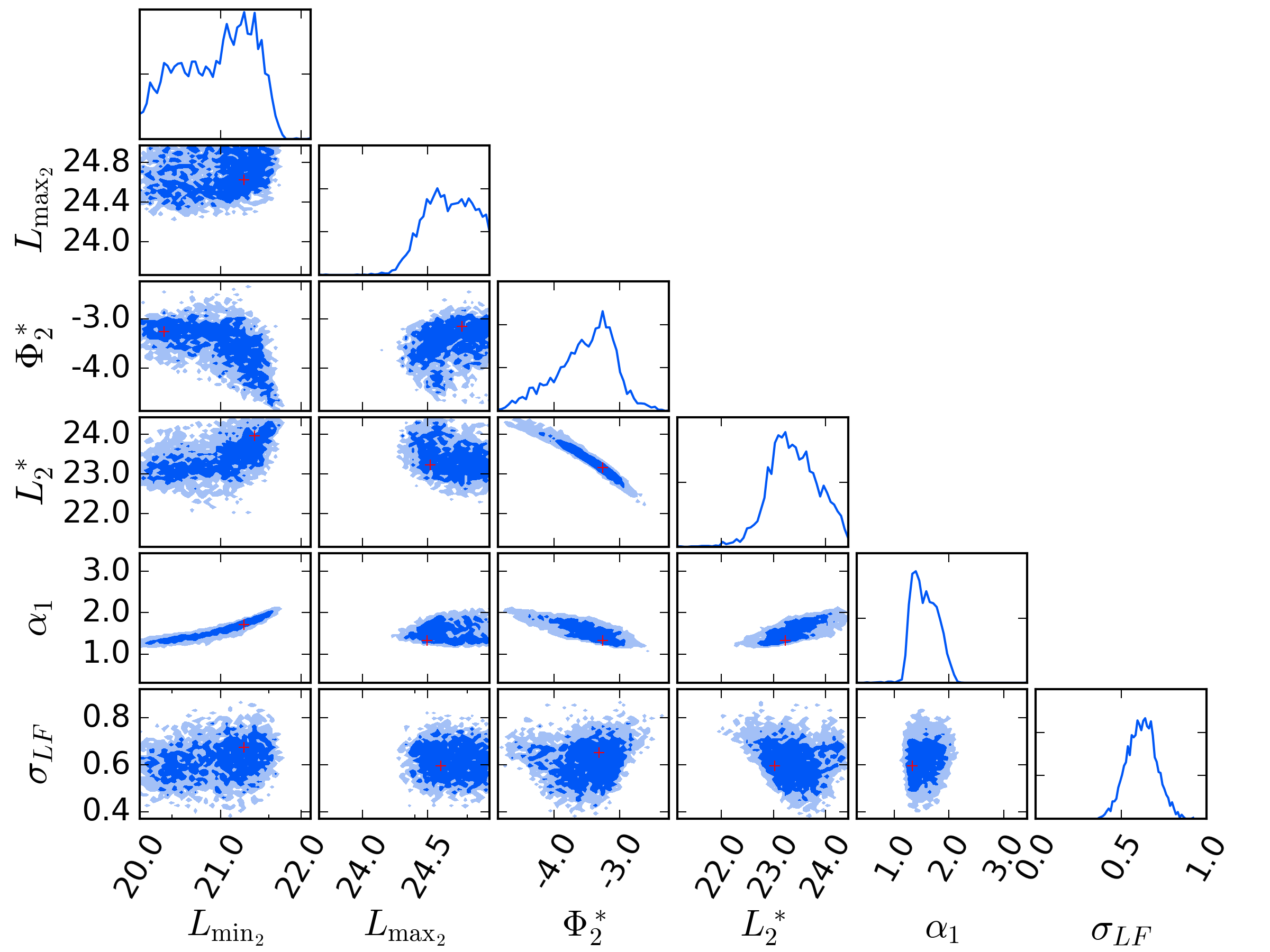}}
\subfloat[Model $C'$;  15 $\mu$Jy]{\includegraphics[width=0.5\textwidth]{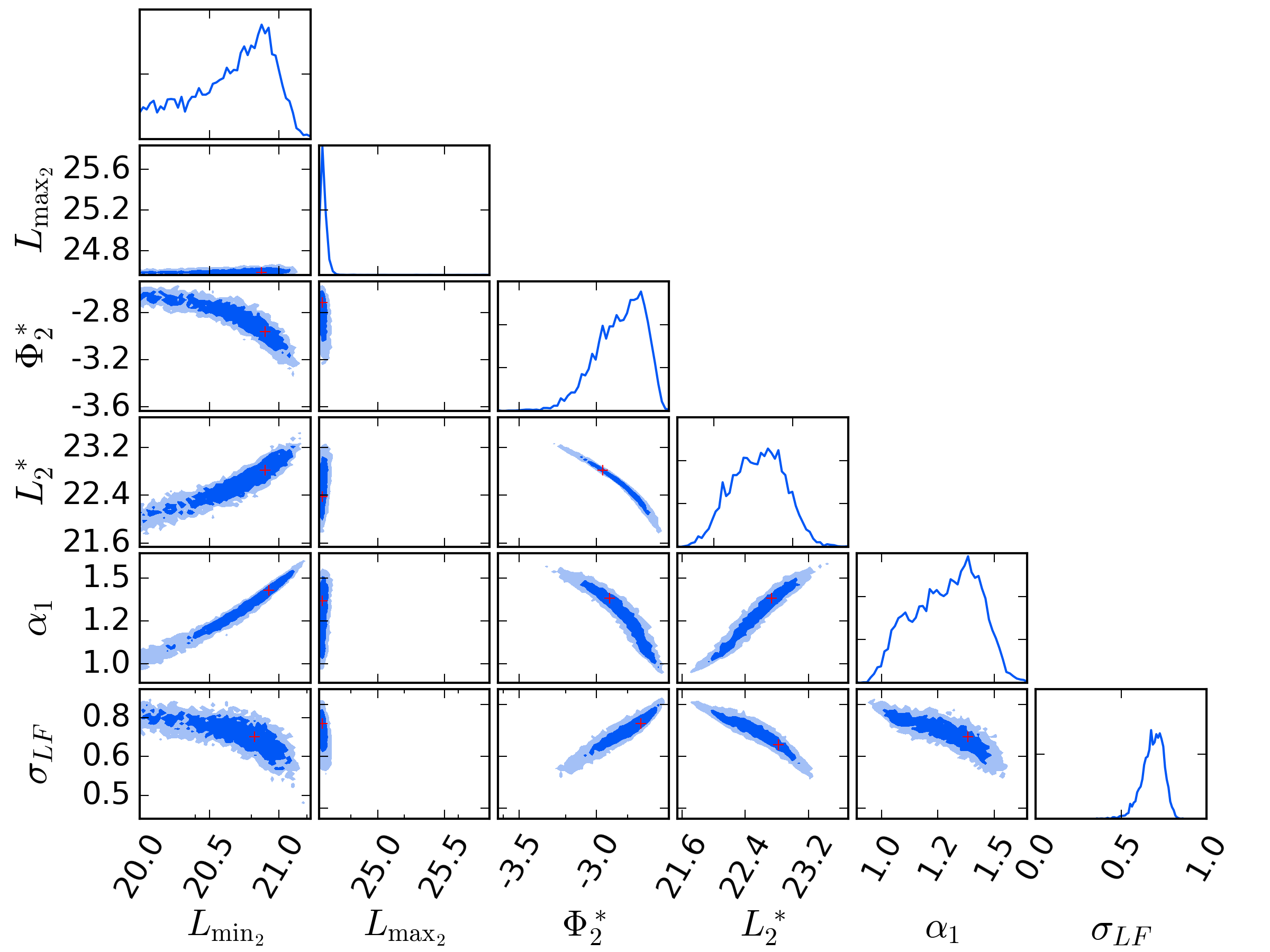}}
\caption{The posterior distributions for the full SKADS sample ($20.5 <\log_{10}[L/(\rm{WHz}^{-1})] < 24.5$) with the noise levels of 150\,$\mu$Jy and 15\,$\mu$Jy using Models $A'$, $B'$ and $C'$. The parameters $L_{\rm{min}_2}$, $L_{\rm{max}_2}$, $\Phi_2^*$ and $L_2^*$ are in logarithmic space ($\log_{10}$). The dark blue and the light blue regions are the 68 per cent and 95 per cent regions. The red cross represents the point with the highest marginalised likelihood in each individual plot.}
\label{fig:tri_skads}
\end{figure*}

\begin{table*}
\centering
\caption{Relative evidence of the different models for each redshift bin in the FIRST data. The reference evidence is the model with the lowest value for each redshift bin, and the winning model has the highest relative evidence (in bold).}
\begin{tabular}{c|ccccccc}
&$\triangle \log_{10}\mathcal{Z}$ & $\triangle \log_{10}\mathcal{Z}$ & $\triangle \log_{10}\mathcal{Z}$ & $\triangle \log_{10}\mathcal{Z}$ & $\triangle \log_{10}\mathcal{Z}$ & $\triangle \log_{10}\mathcal{Z}$ & $\triangle \log_{10}\mathcal{Z}$ \\

\hline

Model& $0.20 < z < 0.45$    & {$0.45 < z < 0.70$} &  {$0.70 < z < 1.00$}  &  {$1.00 < z < 1.30$}  & {$1.30 < z < 1.60$} & $1.60 < z < 1.85$ & $1.85 < z < 2.15$\\
\hline
A & ${4.4 \pm 0.20}$   & ${0.0 \pm 0.00}$   & ${2.0 \pm 0.21}$   & ${10.1 \pm 0.22}$  & ${0.1 \pm 0.23}$  & ${0.0 \pm 0.00}$ & ${0.0 \pm 0.00}$\\
B & $\tb{5.6 \pm 0.20}$&$\tb{0.7 \pm 0.19}$ &$\tb{2.7 \pm 0.21}$&$\tb{11.9 \pm 0.22}$&$\tb{0.8 \pm 0.23}$&$\tb{3.0 \pm 0.22}$& $\tb{2.0 \pm 0.22}$\\
C & ${0.0 \pm 0.00}$   & ${0.6 \pm 0.19}$   & ${0.0 \pm 0.00}$   & ${0.0 \pm 0.00}$   & ${0.0 \pm 0.00}$  & ${1.9 \pm 0.22}$ & ${1.9 \pm 0.25}$\\

\hline
 \hline
\end{tabular}
\label{table:evidence}
\end{table*}

\begin{figure*}
    \centering
    \begin{sideways}
    \includegraphics[width=0.94\textheight]{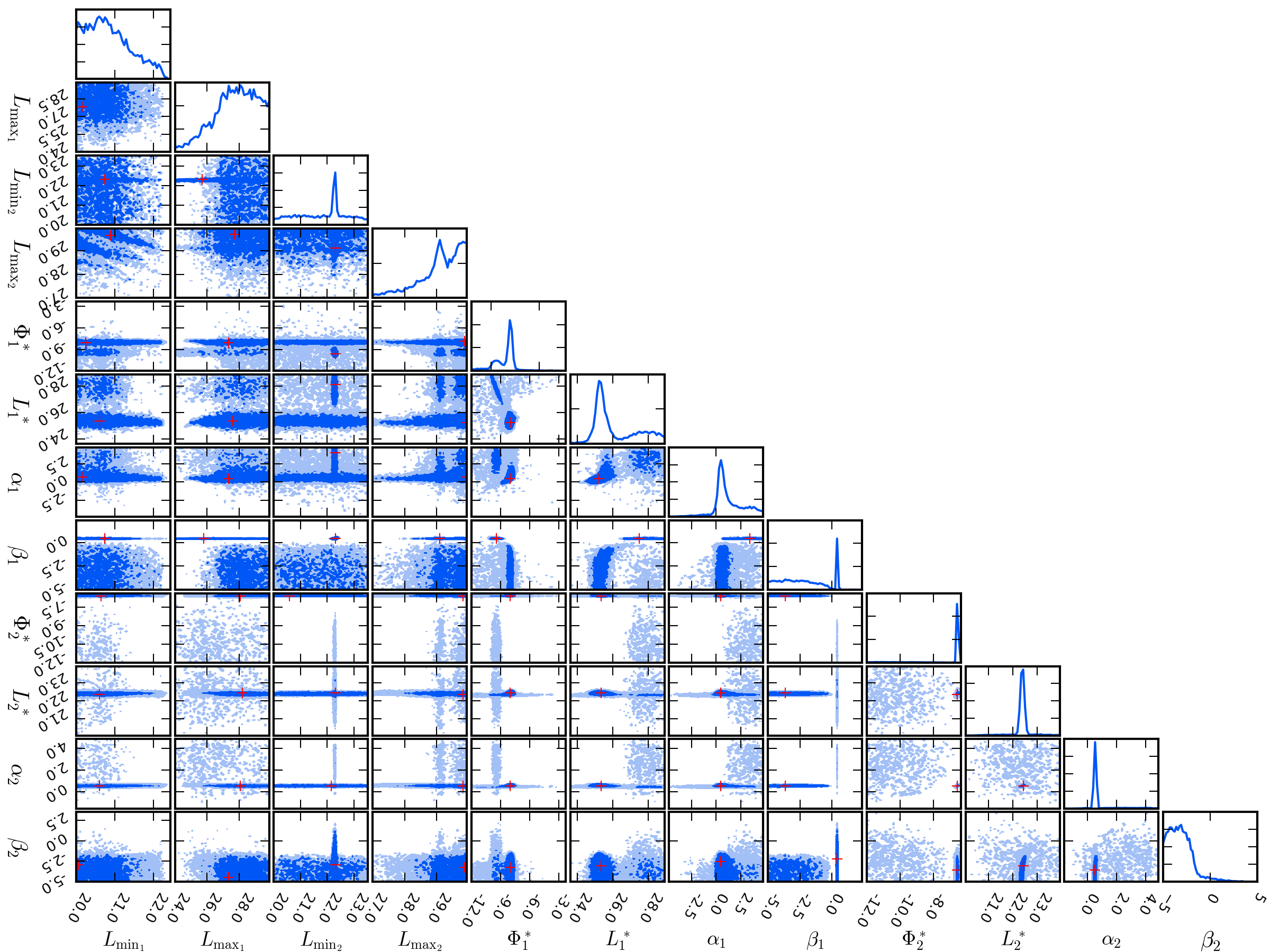}
    \end{sideways}
    \caption{The posterior distributions for the winning model -- model B, the double power-law -- in the lowest redshift bin ($0.2 < z <0.45$). The parameters $L_{\rm{min}_{1,2}}$, $L_{\rm{max}_{1,2}}$, $L_{1,2}*$ and $\Phi_{1,2}^*$ are presented in logarithmic space ($\log_{10}$). The dark blue and the light blue regions are the 68 per cent and 95 per cent regions. The red cross represents the point with the highest marginalised likelihood in each individual plot.}
    \label{fig:vol_post}
\end{figure*}

\section{Results} \label{sec:results}
Having now illustrated the effectiveness of the \textsc{bayestack} algorithm, we apply it to the observational data from SDSS and FIRST. We apply the technique for all three models to each of the redshift bins shown in Table~\ref{table:slices}.

For the high signal-to-noise (detected) flux-densities we can calculate the luminosity function directly by converting flux density to luminosity (neglecting noise) and binning the number of sources in luminosity. 
Although the source populations are volume-limited in the optical (i.e~no brightness cutoff), in the radio some of the sources might not be above the radio flux-density threshold if they are placed at the highest redshift for a given bin. Therefore, we need to apply the $1/V_{\rm{max}}$ correction: the spectral RLF of $N$ sources in a logarithmic bin of width $\Delta m$, using the $1/V_{\rm{max}}$ method \citep{Schmidt-1968} is given by,
\begin{equation}
\Phi(L_\nu) = \frac{1}{\Delta m}\sum^N_{i=1}\left(\frac{1}{V_{\rm{max}}}\right)_i, 
\end{equation}
with an uncertainty
\begin{equation}
\sigma(\Phi) = \frac{1}{\Delta m}\left[ \sum^N_{i=1}\left(\frac{1}{V_{\rm{max}}}\right)_i^2 \right]^{1/2}, 
\end{equation}
where $V_{\rm{max}}$ is the maximum comoving volume at which the source is detected.

\subsection{Quasars at $0.2<z<0.45$} \label{sec:lowz}
We start with the lowest redshift sample because it allows a direct comparison to the work of \cite{Kellermann-2016}.

We use the Bayesian technique to fit the three RLF models to all the sources (radio detected and undetected) from a volume-limited sample defined by $M_i < -23$ at $0.2 < z < 0.45$. The first column of \T{table:evidence} shows the relative evidence of the fit to the models. From the relative evidence we conclude that the data significantly prefer Model B, which consists of a double power-law for the luminous sources and a second double power-law for the low-luminosity and undetected sources.

\F{fig:vol_post} shows the posterior distributions for the winning Model B. Like with the SKADS sources, the boundary parameters $L_{\rm{max}_{1,2}}$ along with $L_{\rm{min}_{1,2}}$  exceed the prior limit and are unconstrained. Note, however, that this has very little impact on the actual observed numbers (and that the uncertainty increases due to noise and/or Poisson fluctuations). The other parameters have well-defined peaks, except for the faint-end slopes $\beta_{1,2}$ which span a large range below 0. 

In Fig.~\ref{fig:LF_multi} we show the optically-selected quasar RLF across the full luminosity and redshift range from our sample. The black circles denote the RLF determined using the $1/V_{\max}$ method, which is only possible for those detected above a certain flux-density threshold (we use $5\sigma$), whereas the lines and shaded regions show the full RLF distribution from the Bayesian modelling. 
Concentrating on the lowest redshift bin (top-left panel of Fig.~\ref{fig:LF_multi}), we find that the number density of radio-bright quasars increases with decreasing radio luminosity in all redshift bins, as expected.

We compare our inferred RLFs for optically-selected quasars with similar RLFs from the literature: \cite{Condon-2013} and \cite{Kellermann-2016}. The optical data are the same volume-limited sample from SDSS's DR7. The radio data are all from the VLA but each sample was observed with different configurations and depths. Our data are from FIRST, which was observed in the `B' configuration, with a resolution of $5^{\prime\prime}$ and rms of 0.15~mJy, corresponding to a detection threshold of 1~mJy. The \citeauthor{Condon-2013} sample is from the NRAO VLA Sky Survey (NVSS; \citealt{Condon-1998}) observed using the compact `D' and `DnC' configurations, with a resolution of $45^{\prime\prime}$ and rms of 0.45~mJy (a detection threshold of 2.4~mJy). \citeauthor{Kellermann-2016} observed a complete sub-sample of these quasars, over a reduced redshift range ($0.2<z<0.3$) at 6\,GHz using the Karl G.~Jansky Very Large Array (JVLA) in the `C' configuration, with a resolution of $3.5^{\prime\prime}$ and rms as low as $6\mu$Jy for the fainter sources. In order to enable a direct comparison to the results in our lowest redshift bin, the 6\,GHz luminosities of the \citeauthor{Kellermann-2016} sources are converted to 1.4~GHz luminosities using a spectral index of $\alpha =0.7$ and their number density is increased by $\log_{10}[\Phi/(\rm{Mpc}^{-3} \rm{mag}^{-1})] = 0.2$  to correct for evolution (\citealt{Condon-2013}) when comparing the RLF over the redshift range $0.2<z<0.3$ with that over $0.2<z<0.45$.

 The RLF above the nominal 5-$\sigma$ threshold for our sample is in good agreement with both \cite{Condon-2013} and \cite{Kellermann-2016} RLFs between radio luminosities $25 < \log_{10}[L_{1.4}/{\rm W\,Hz}^{-1}] < 26$ but is less consistent with the \cite{Condon-2013} RLF towards the low-luminosity end of where we have direct detections ($23.6 < \log_{10}[L_{1.4}/{\rm W\,Hz}^{-1}] < 25$).
 Furthermore, our RLF has large uncertainties above $\log_{10}[L_{1.4}/{\rm W\,Hz}^{-1}] \sim 26$. These are both likely due to the fact that only 7 of the 26 sources observed in NVSS are compact (\citealt{Condon-2013}) and the rest are extended sources that have emission resolved out by FIRST (hence the sources occupy lower luminosity bins below $\log_{10}[L_{1.4}/{\rm W\,Hz}^{-1}] \sim 26$), and as such lead to the discrepancy with the \cite{Condon-2013} study and reduce the numbers in the highest luminosity bins.  
 
 Each of the RLFs in this redshift bin also show a flattening in the number density between $\log_{10}[L_{1.4}/{\rm W\,Hz}^{-1}]  \approx 25.5$ and $\log_{10}[L_{1.4}/{\rm W\,Hz}^{-1}]  \approx 24$. Below $\log_{10}[L_{1.4}/{\rm W\,Hz}^{-1}] =24.4$ our RLF is higher than that of \citeauthor{Condon-2013} but still in good agreement with \citeauthor{Kellermann-2016}. The difference between our RLF and that from \citeauthor{Condon-2013} is most probably due to the difference in resolution of the radio data, which results in sources moving into lower-luminosity bins due to some emission being resolved out.

 Given the likely underestimation of extended emission using the FIRST survey, we use the \citeauthor{Condon-2013} flux-densities for sources found in both NVSS and FIRST in the RLF fit (shown in \F{fig:vol_post} and \F{fig:LF_multi}). However, we note that extended emission may still be resolved out for sources below the flux-density limit but we have no way of estimating this. Although we could potentially use the NVSS data here, we would then have to deal with confusion issues due to the larger synthesised beam. We therefore continue to use the FIRST data, but the issue of extended emission should be borne in mind.

The reconstruction of the RLF below the detection threshold continues to follow the slope (of the double power-law) established from $\log_{10}[L_{1.4}/{\rm W\,Hz}^{-1}] \ge 24.4$, dropping at $\log_{10}[L_{1.4}/{\rm W\,Hz}^{-1}] \approx 22.4$. This therefore measures the RLF $\sim 2$ orders of magnitude below $\log_{10}[L_{1.4}/{\rm W\,Hz}^{-1}] = 23.4$ ($=5\sigma_n $). The steep drop-off in the RLF at $\log_{10}[L_{1.4}/{\rm W\,Hz}^{-1}] < 22$ is due to the optical limit of the quasar sample, meaning that there are no optically-selected quasars with $i < 19.1$ contributing to this part of the RLF.

Comparing the reconstructed RLF to the \cite{Kellermann-2016} individually-observed sources, we find that the two measurements are in complete agreement.

\begin{figure*}
\centering
\includegraphics[width=0.9\textwidth]{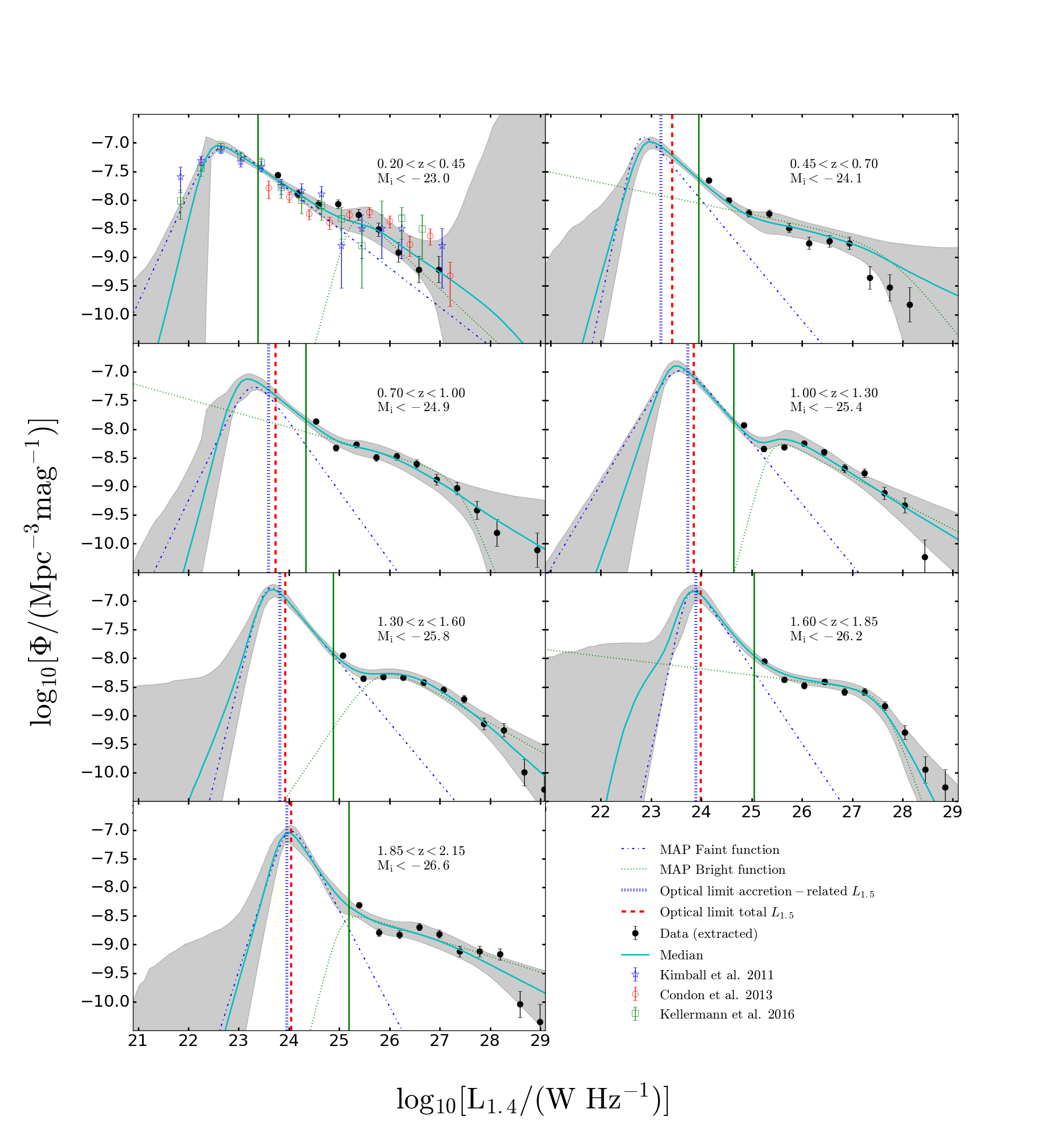}
\caption{The optically-selected quasar radio luminosity function and its evolution with redshift. The black dots are the $1/V_{\rm{max}}$ RLFs from sources above $5\sigma_n$. The blue unfilled stars, red unfilled circles and the green unfilled squares respectively represent the RLFs from \protect\cite{Kimball-2011}, \protect\cite{Condon-2013} and \protect\cite{Kellermann-2016}. The cyan curves represent the RLFs reconstructed from the median parameters of the winning model (\T{table:evidence}) in each redshift bin. The blue dashed-dotted and green dotted lines are the faint and bright functions with their MAP parameters respectively. The grey region represents the 95-per-cent confidence interval of the distribution of reconstructions of models in the posterior. The green, vertical line is $5\sigma_n$ computed using the mean redshift for each redshift bin. The blue dotted and red dashed lines are an estimate of the radio-luminosity limit that corresponds to the optical limit, assuming optical--radio correlations for quasars \citep{White-2017} that are based on accretion-related radio-emission and total radio-emission, respectively.}
\label{fig:LF_multi}
\end{figure*}

\begin{table*}
\centering
\caption{The MAP posterior parameters for the double power-law -- the winning model -- for the quasar RLF, in each of the redshift bins and their $2\sigma$. The units of the parameters are as shown in \T{table:prior}.}
\begin{tabular}{l|rrrrrrr}
\hline
                Parameter & $0.2 < z < 0.45$ & $0.45 < z < 0.7$ & $0.7 < z < 1.0$ & $1.0 < z < 1.3$ & $1.3 < z < 1.6$ & $1.6 < z < 1.85$ & $1.85 < z < 2.15$ \\ \hline 
$\log_{10}[L_{\rm{min_1}}]$ &$ 20.83_{-0.79}^{+1.27}$ & $ 20.69_{-0.65}^{+1.69}$ & $ 21.94_{-1.87}^{+0.85}$ & $ 20.76_{-0.71}^{+1.97}$ & $ 20.45_{-0.40}^{+2.68}$ & $ 22.00_{-1.90}^{+1.38}$ & $ 22.38_{-2.35}^{+0.40}$ \\ [3pt]
$\log_{10}[L_{\rm{max_1}}]$ &$ 28.75_{-3.82}^{+1.14}$ & $ 24.18_{-0.91}^{+5.65}$ & $ 24.69_{-0.07}^{+3.25}$ & $ 29.44_{-3.98}^{+0.43}$ & $ 29.35_{-3.92}^{+0.52}$ & $ 26.30_{-0.85}^{+3.57}$ & $ 25.92_{-0.29}^{+3.94}$ \\ [3pt]
$\log_{10}[L_{\rm{min_2}}]$ &$ 20.60_{-0.44}^{+2.85}$ & $ 23.43_{-3.29}^{+0.94}$ & $ 22.69_{-2.54}^{+1.67}$ & $ 21.26_{-3.05}^{+3.89}$ & $ 18.63_{-0.39}^{+6.43}$ & $ 22.31_{-2.11}^{+2.54}$ & $ 22.71_{-2.51}^{+2.55}$ \\ [3pt]
$\log_{10}[L_{\rm{max_2}}]$ &$ 29.65_{-2.18}^{+0.32}$ & $ 28.91_{-1.36}^{+1.05}$ & $ 29.03_{-0.91}^{+0.94}$ & $ 29.76_{-1.49}^{+0.22}$ & $ 29.55_{-1.49}^{+0.42}$ & $ 28.12_{-0.09}^{+1.83}$ & $ 29.92_{-1.26}^{+0.07}$ \\ [3pt]
    $\log_{10}[\Phi_1^*]$ &$ -8.01_{-3.50}^{+1.57}$ & $ -9.02_{-0.16}^{+0.68}$ & $ -8.79_{-0.35}^{+0.72}$ & $ -8.10_{-0.12}^{+0.26}$ & $ -8.03_{-0.26}^{+0.10}$ & $ -8.58_{-0.18}^{+0.42}$ & $ -8.39_{-0.35}^{+0.16}$ \\ [3pt]
       $\log_{10}[L_1^*]$ &$ 25.25_{-0.55}^{+3.52}$ & $ 27.74_{-2.72}^{+0.08}$ & $ 27.36_{-2.39}^{+0.37}$ & $ 25.38_{-0.13}^{+0.70}$ & $ 25.78_{-0.08}^{+1.41}$ & $ 27.61_{-0.82}^{+0.22}$ & $ 25.04_{-0.01}^{+2.29}$ \\ [3pt]
               $\alpha_1$ &$  0.82_{-2.45}^{+3.80}$ & $  1.16_{-1.10}^{+2.74}$ & $  2.27_{-2.09}^{+1.21}$ & $  0.46_{-0.11}^{+0.39}$ & $  0.50_{-0.08}^{+0.70}$ & $  2.46_{-1.56}^{+1.00}$ & $  0.27_{-0.04}^{+0.70}$ \\ [3pt]
                $\beta_1$ &$ -3.26_{-1.60}^{+3.79}$ & $  0.18_{-4.70}^{+0.01}$ & $  0.24_{-4.90}^{+0.10}$ & $ -3.30_{-1.45}^{+2.84}$ & $ -1.31_{-1.19}^{+1.30}$ & $  0.11_{-0.56}^{+0.11}$ & $ -3.44_{-1.29}^{+3.53}$ \\ [3pt]
    $\log_{10}[\Phi_2^*]$ &$ -6.73_{-4.91}^{+0.04}$ & $ -6.94_{-0.16}^{+0.00}$ & $ -6.97_{-0.12}^{+0.15}$ & $ -6.68_{-0.02}^{+0.11}$ & $ -6.53_{-0.09}^{+0.07}$ & $ -6.55_{-0.12}^{+0.10}$ & $ -6.73_{-0.19}^{+0.09}$ \\ [3pt]
       $\log_{10}[L_2^*]$ &$ 22.55_{-1.36}^{+0.78}$ & $ 22.74_{-0.14}^{+0.20}$ & $ 23.25_{-0.42}^{+0.22}$ & $ 23.54_{-0.27}^{+0.02}$ & $ 23.53_{-0.11}^{+0.26}$ & $ 23.76_{-0.20}^{+0.23}$ & $ 24.00_{-0.32}^{+0.07}$ \\ [3pt]
               $\alpha_2$ &$  0.68_{-0.24}^{+3.37}$ & $  1.05_{-0.40}^{+0.06}$ & $  1.21_{-0.65}^{+0.54}$ & $  1.07_{-0.22}^{+0.14}$ & $  1.05_{-0.17}^{+0.47}$ & $  1.31_{-0.40}^{+0.47}$ & $  1.67_{-0.80}^{+0.26}$ \\ [3pt]
                $\beta_2$ &$ -1.92_{-2.97}^{+1.92}$ & $ -4.21_{-0.69}^{+2.67}$ & $ -1.82_{-3.09}^{+0.42}$ & $ -1.48_{-3.31}^{+0.06}$ & $ -3.56_{-1.33}^{+2.22}$ & $ -4.05_{-0.85}^{+2.67}$ & $ -2.75_{-2.17}^{+0.96}$ \\ [3pt]

\hline
 \hline
\end{tabular}
\label{table:params}
\end{table*}
  
\subsection{Higher-redshift bins}

In Sec.~\ref{sec:lowz} we demonstrated that the technique is able to reconstruct the RLF below the detection threshold in the lowest-$z$ sample, for which there are deeper radio data. In this section we present the result using our algorithm and the three models describing the RLF to the higher redshift bins. The relative evidence of the models for each redshift bin are shown in \T{table:evidence}. The data prefer Model B (a double power-law for the low-luminosity sources) for all the redshift bins. The posterior distributions for the winning models are shown in the online version of the paper.
 
The optically-selected quasar RLF mirrors the general shape seen in the lowest redshift bin over all redshifts. In all cases we see that the bright-end of the RLF increases steeply as the radio luminosity decreases towards $L_1^* \sim 10^{25}$\,W\,Hz$^{-1}$ and then turns over. Just below this luminosity we see the second (faint-end) double power law starting to dominate the RLF, where we find a steep increase as the radio luminosity decreases towards $\log_{10}[L_{1.4}/{\rm W\,Hz}^{-1}] \sim {23}$. Our reconstructed RLF also follows the 1/$V_{\rm{max}}$ points very well where we are able to measure them.
 
This flattening of the bright-end of the RLF and subsequent increase below $\log_{10}[L_{1.4}/{\rm W\,Hz}^{-1}] < 26$ is also observed in  optically-selected quasar RLFs studies \citep[e.g.][]{Condon-2013,Kellermann-2016,Hwang-2018}. A similar flattening is also observed in the RLF of other optically-selected-AGN samples \citep[e.g.][]{Rush-1996,Padovani-2015}. A clear change of the slope in the number density is also observed in radio-selected AGN RLFs (e.g \citealt{Willott-2001}, \citealt{Smolcic-2009}, \citealt{McAlpine-2013}). Indeed, our fitted values for $L_1^*$ are in good agreement with the RLF derived using the deep VLA-3\,GHz survey from \cite{Smolcic-2017}.

At radio luminosities below where the flattening takes place, the reconstructed RLF steeply increases towards lower luminosities, with a slope established above $5\sigma_n$ for all redshift bins. The preferred model for all redshift bins is Model B (Table~\ref{table:evidence}, the double power-law). With this model, the RLF in all redshift bins has a peak at $L_2^*$, dropping below $L_2^*$. This fall-off is due to the hard absolute magnitude cut-off in the parent sample, and essentially means that there is no significant evidence for any radio continuum emission from our quasar sample below $L_2^*$.

\section{Discussion} \label{sec:discusion}	
The definition of radio-loudness varies in the literature, as some objects can be classified as `radio-quiet' in one definition and `radio-loud' in another, e.g. either by considering the ratio of optical to radio emission \citep[e.g.][]{Kellermann-1989} or by just using a radio lumninosity threshold \citep[e.g.][]{Miller-1990}. In this paper, we do not explicitly classify our quasars as radio-loud or radio-quiet, instead using the shape of the RLF to infer where these populations dominate. In all of our RLFs (\F{fig:LF_multi}) there is a clear change in behaviour at or around $\log_{10}[L_{1.4}/{\rm W\,Hz}^{-1}] \sim 25$. We define the `radio-loud' population as the quasars that are described by a bright-end double power-law (parameters with subscript `1' in the modelling). The faint end (radio-quiet quasars) is parameterised by the power-law, double power-law or log-normal function. For this study all redshift bins had the double power-law as the winning model (\T{table:evidence}).  

\subsection{Radio-loud quasars}

The radio emission from radio-loud quasars are powered by processes associated with the accretion on to the central supermassive black hole. Falling within the AGN orientation-based unification model \citep[e.g.][]{Barthel-1989, Antonucci-1993, Urry-1995}, these radio-loud objects have been shown to require a supermassive black-hole of mass $M_{\rm BH} > 10^8$\,M$_{\odot}$ \citep{McLureJarvis2004}, whereas their radio-quiet counterparts can have lower-mass black holes.
By integrating under the two double power-law models, representing the bright- and faint-end of the RLF, we find that the radio-loud fraction of quasars make up $\approx$\,10~per~cent of the total quasar population in our sample at $z> 0.7$ (\T{table:slices}). However, we find that the radio-loud fraction drops to $\approx$\,7 and 4~per~cent of the total quasars in the two lowest redshift bins (Table~\ref{table:slices}). This lower fraction of radio-loud quasars towards lower redshifts reflects the fact that we have a much fainter optical magnitude limit at low redshift, and if radio-loudness is linked to the combination of accretion rate and black-hole mass, then lower-optical luminosity quasars are more likely to be radio quiet. These fractions are in line with previous studies of radio-loud and radio-quiet quasars with a variety of classification schemes \citep[e.g.][]{White-2007,Cirasuolo-2005,Balokovic-2012}.

However, one of the differences is that we actually find a much more pronounced flattening than the studies based purely on radio-selected samples \citep[e.g.][]{Willott-2001,Smolcic-2009,McAlpine-2013}. One reason for this could be that there is a real difference in the physical properties that generate radio emission in optically-selected quasars compared to the more-general population of radio-selected AGN. We also cannot rule out the possibility of the optical selection creating a bias in the RLF that artificially flattens, or decreases, the bright-end of the RLF below $\log_{10}[L_{1.4}/{\rm W\,Hz}^{-1}] \sim 26$. However, we have been conservative with our optical selection, ensuring that the quasar sample is complete across the full width of all redshift bins. We cannot rule out incompleteness due to the colour selection within the SDSS sample, but we would not expect this to have a significant effect in individual, relatively narrow, redshift bins. A possible explanation could be due to our sample becoming incomplete in terms of the RLF based on the optically-selected sample. This could arise if there is a correlation between the optical emission in these quasars and their radio emission. 

Several authors have investigated the link between optical emission and radio emission from quasars \citep[e.g.][]{Serjeant-1998,White-2007,White-2017}, finding evidence for a correlation. However, one has to be careful when measuring correlations between flux-limited samples. Therefore, in Fig.~\ref{fig:LF_multi} we show the radio luminosity where we expect the optical flux limit to start imposing incompleteness on the RLF, based on the absolute magnitude limits shown in Table~\ref{table:slices}. For this we use the relation between optical luminosity and the star-formation subtracted radio luminosity found by \cite{White-2017}, from their radio-quiet quasar sample at $z\sim 1$. We also show the radio luminosity limit based on the \cite{White-2017} optical luminosity versus total radio-luminosity, for completeness. One can see that the radio luminosity at which the optical selection may lead to incompleteness in the RLF is just above the radio luminosity at which the drop in the RLF occurs. This supports the argument that the drop/turnover at low luminosities is caused by lack of fainter optical quasars. However, there is significant scatter in the \cite{White-2017} optical-radio correlation of around 1 order of magnitude in radio luminosity for a given optical luminosity. Therefore, it is certainly possible that some of the flattening could arise from incompleteness in the RLF due to the optical magnitude limit. To test this we increased the optical magnitude limit for our sample in each redshift bin in order to check if the flattening or downturn becomes more prominent. In all bins the turnover (i.e. the value of $\beta_1$) became more prominent. We therefore suggest that at least some of the flattening is due to incompleteness introduced by the optical magnitude limit of the parent sample, although we note that the uncertainties increase due using a smaller quasar sample when a higher optical-luminosity threshold is imposed..

\subsection{Radio-quiet quasars}
This population makes up about 92 per cent (\T{table:slices}) of the quasar population in our sample, but the origin of the radio emission is not well understood. 
Our reconstructed RLFs increase steeply below $\log_{10}[L_{1.4}/{\rm W\,Hz}^{-1}] \sim {24.5}$. This steepening could be attributed to an increasing contribution from SF in the host galaxy (e.g.~\citealt{Terlevich-1987}, \citeyear{Terlevich-1992}; \citealt{Padovani-2011}; \citealt{Kimball-2011}; \citealt{Bonzini-2013}; \citealt{Condon-2013}; \citealt{Kellermann-2016}; \citealt{Stacey-2018}; \citealt{Gurkan-2018}) or is AGN-related with a different scaling relation or different emission associated with the AGN 
(\citealt{Herrera_Ruiz-2016}, \citealt{Zakamska-2016}, \citealt{White-2015}, \citeyear{White-2017}, \citealt{Hartley-2019}) compared to their radio-loud counterparts. Although, we note that the steepening is significantly less pronounced in the two lowest-redshift bins, which may indicate that the optical magnitude limit may play a role in creating an artificially-steepening slope in the observed RLF. In such a case, the distinction between radio-loud and radio-quiet would become more difficult, with evidence that the population has a more continuous distribution 
\citep[e.g.][]{Lacy-2001,Gurkan-2019}.

\cite{Kimball-2011} suggested that the `bump' observed at $\log_{10}[L_{1.4}/{\rm W\,Hz}^{-1}] \approx 22.7$ in their low-$z$ volume-limited sample (\F{fig:LF_multi}) corresponds to star-forming galaxies.  \cite{Kellermann-2016} tested their hypothesis by using mid-infrared data from WISE to search for a correlation between the 22 $\mu$m and 6-GHz flux-densities, which is a characteristic of the radio--far-infrared correlation. However, they found no strong correlation and so suggest that the 22\,$\mu$m fluxes do not only measure SF but can also be contaminated by warm dust heated by the AGN (\citealt{Polletta-2010}). \cite{Coziol-2017} tested the SF hypothesis by also matching the \citeauthor{Kellermann-2016} sources against WISE. They found counterparts for all but 7 sources, created a new diagnostic plane based on WISE colours (\citealt{Coziol-2015}), and found that: (i) there is no separation between the radio-quiet and radio-loud quasars in the colour distribution, and (ii) the majority of the \citealt{Kellermann-2016} quasars (and our lowest $z$ sample) have low star-formation rates. 

\cite{White-2015} used deep optical and near-infrared data to identify a sample of quasars across a range of redshifts, and conducted a stacking experiment using deep VLA 1.4\,GHz data. Comparing their quasar sample with multiple galaxy samples (matched by stellar mass, having assumed the black-hole mass of the quasar), they provided evidence that the radio emission from these quasars -- which lie at much higher redshift but cover similar  optical luminosities as the \cite{Kellermann-2016} sample -- predominantly arises from accretion-related activity. Furthermore, by comparing the star-formation rates using far-infrared data of a randomly selected subset of a volume-limited quasar sample at $0.9<z<1.1$, \cite{White-2017} showed that the radio emission from star formation is sub-dominant. 

The only evidence in our modelled RLFs for star-formation contributing to the radio emission in quasars comes from the observed strong steepening of the RLF towards low luminosities, below the nominal 5$\sigma$ detection threshold at $z> 0.7$. However, where our optically-selected quasar sample contains the lowest-luminosity quasars ($z<0.7$), the evidence for this steepening is weaker. On the other hand, comparing the observed upturn in the quasar RLF with the star-forming galaxy RLF at $z>0.8$ from \cite{Novak-2018}, we find that the steepening occurs at approximately the same radio luminosity that the star-forming galaxies dominate over AGN in radio-selected surveys. This strengthens the suggestion that star formation plays an important role at these low radio luminosities.
Indeed, this was used as evidence in favour of the star-formation becoming the dominant contribution to the radio luminosity in this regime by \cite{Kimball-2011} and \cite{Condon-2013}. Nevertheless, it is clear that the RLF is a relatively blunt tool for disentangling the dominant contribution to the low-luminosity radio emission in quasars. A more productive route may be to explore the bivariate optical and radio luminosity function for quasars \citep[e.g.][]{Singal-2011}, where the optical selection is naturally accounted for and models that link the optical and radio emission could be incorporated. 

A more direct method would be to use high-resolution radio data that can resolve any star formation on the scale of the host galaxy. The VLA has the potential to do this, but would need to move towards a frequency of 6\,GHz to achieve the required resolution of $\sim 0.5\arcsec$ for the vast majority of quasars that lie at $z>0.5$. Given the typical spectral index of the synchrotron radiation from both star-formation and AGN-associated emission of  $\alpha \sim 0.7$, this would then require  longer integration times and may also suffer from contamination from free-free emission, making the results more difficult to interpret. eMERLIN has the potential to carry out similar resolution studies at lower frequencies \citep[e.g.][]{Guidetti-2013,Radcliffe-2018,Jarvis-2019}. In the future, the Square Kilometre Array \citep[e.g.][]{Jarvis-2004,Smolcic-2015,Mcalpine-2015} would be able to carry out high-resolution studies to much deeper levels at a range of frequencies and thus help make great strides in our understanding of the dominant radio emission mechanism in radio-quiet quasars.

\section{Conclusions} \label{sec:conclusion}
\begin{enumerate}[i]
\item We have built on the work of \cite{Roseboom_Best-2014} and \cite{Zwart-2015} by fitting directly quasar radio luminosity functions below the radio detection threshold using a Bayesian stacking approach (\textsc{bayestack}). We tested the technique by fitting three models to mock SKADS simulation catalogues (\citealt{Wilman-2008}; \citeyear{Wilman-2010}), with random Gaussian noise of 150 $\mu$Jy added. We successfully recovered the SKADS RLF over three orders of magnitude below the $5\sigma$ detection threshold. We ran further tests using mock catalogues with 15 $\mu$Jy Gaussian noise and as expected reconstructed a better-constrained RLF with respect to the true SKADS RLF.

\item We used FIRST radio flux-densities extracted at the positions of optical quasars from a uniformly-selected (homogeneous) sample of SDSS DR7 divided into seven volume-limited redshift bins. We parameterised the high-luminosity RLF using a double power-law. For our lowest-$z$ sample we found that the $1/V_{\rm{max}}$ and double power-law RLF for luminous sources is in agreement with that from \citeauthor{Kellermann-2016}, but is marginally inconsistent with that of \citeauthor{Condon-2013} at the luminous and faint ends of the detected RLF. 
Some of the difference at the faint end is likely due to the different resolution of NVSS and FIRST. In the other redshift bins, we find that each of the bright ends of the RLFs, which broadly represent radio-loud quasars, are  well described by a double power-law. This double power-law generally flattens towards low luminosities. A similar drop/flattening is observed for extremely-red quasars \citep{Hwang-2018} and in AGN RLFs (e.g \citealt{Smolcic-2009}). We suggest that some of this flattening could also be due to the optical flux limit of the sample reducing the number of quasars that could contribute radio data to these radio luminosities, although this would need to be tested thoroughly with a deeper optical selection or by considering a bivariate model of the optical and radio luminosity functions.

\item With \textsc{bayestack} we probe the RLF down to 2 orders of magnitude below the detection threshold of FIRST (1~mJy). The difference in how deep we can probe compared to the simulation is related to the lack of low luminosity radio sources in the sample as compared to SKADS.
At low redshift ($z<0.7$) we see a continuous distribution from the bright to the faint end of the RLF, whereas at $z>0.7$, the RLF steeply increases towards fainter luminosities $\log_{10}[L_{1.4}/{\rm W\,Hz}^{-1}] < {24.5}$. This could be due to the source population changing or due to the biased flattening of the RLF because of the optical flux limit described previously. We note, however, that the steep increase coincides with the measured steepening in the RLF from radio-selected samples of star-forming galaxies \citep[e.g.][]{Novak-2018}. In order to resolve whether this steepening is indeed due to star formation, higher-resolution radio imaging would be ideal, in order to resolve the radio emission from star formation in the host galaxy.

\item Finally, the RLF peaks between $\log_{10}[L_{1.4}/{\rm W\,Hz}^{-1}] \sim {22.5}$ and $\log_{10}[L_{1.4}/{\rm W\,Hz}^{-1}] \sim {24}$ (depending on the redshift) and drops rather abruptly after that. This is due to the parent sample containing no quasars that are generating radio emission below this luminosity.

\end{enumerate}

\subsection*{Acknowledgments}
We thank the referee for the helpful comments that have contributed to improve this paper.
EM, MGS, MJJ and SVW acknowledge support from the South African Radio Astronomy Observatory (SARAO). EM and MGS also acknowledge support from the National Research Foundation (Grant No.~84156). JZ is thankful for a South African Square Kilometre Array Research Fellowship. We are grateful for valuable contributions from Matt Prescott, Imogen Whittam, Margherita Molaro, Jo\'se Fonseca and Brandon  Engelbrecht. We would like to acknowledge the computational resources of the Centre for High Performance Computing. 
\section*{Appendix}
\begin{appendix}
\renewcommand{\thefigure}{A\arabic{figure}}
\setcounter{figure}{0}

\F{fig:tri_1} show the 1-D and 2-D posterior distributions for all of the winning models for each redshift bin. The 1-D posterior distribution is the marginalization of each parameter shown at the end of each row. The parameters have well-defined peaks, except for the boundary parameters ($L_{\rm{min}_1}$, $L_{\rm{max}_1}$, $L_{\rm{min}_2}$ and $L_{\rm{max}_2}$) and the second slope $\beta_2$ (the faint-end slope for the faint-quasar function in Models A and B), which are not well constrained. The upper limit of the fitted $L_{\rm{max}_1}$ is unconstrained or (even) truncated, but this does not significantly affect the fit in our areas of interest.






\begin{figure*}
\centering
\ContinuedFloat
\begin{sideways}
\includegraphics[width=0.92\textheight]{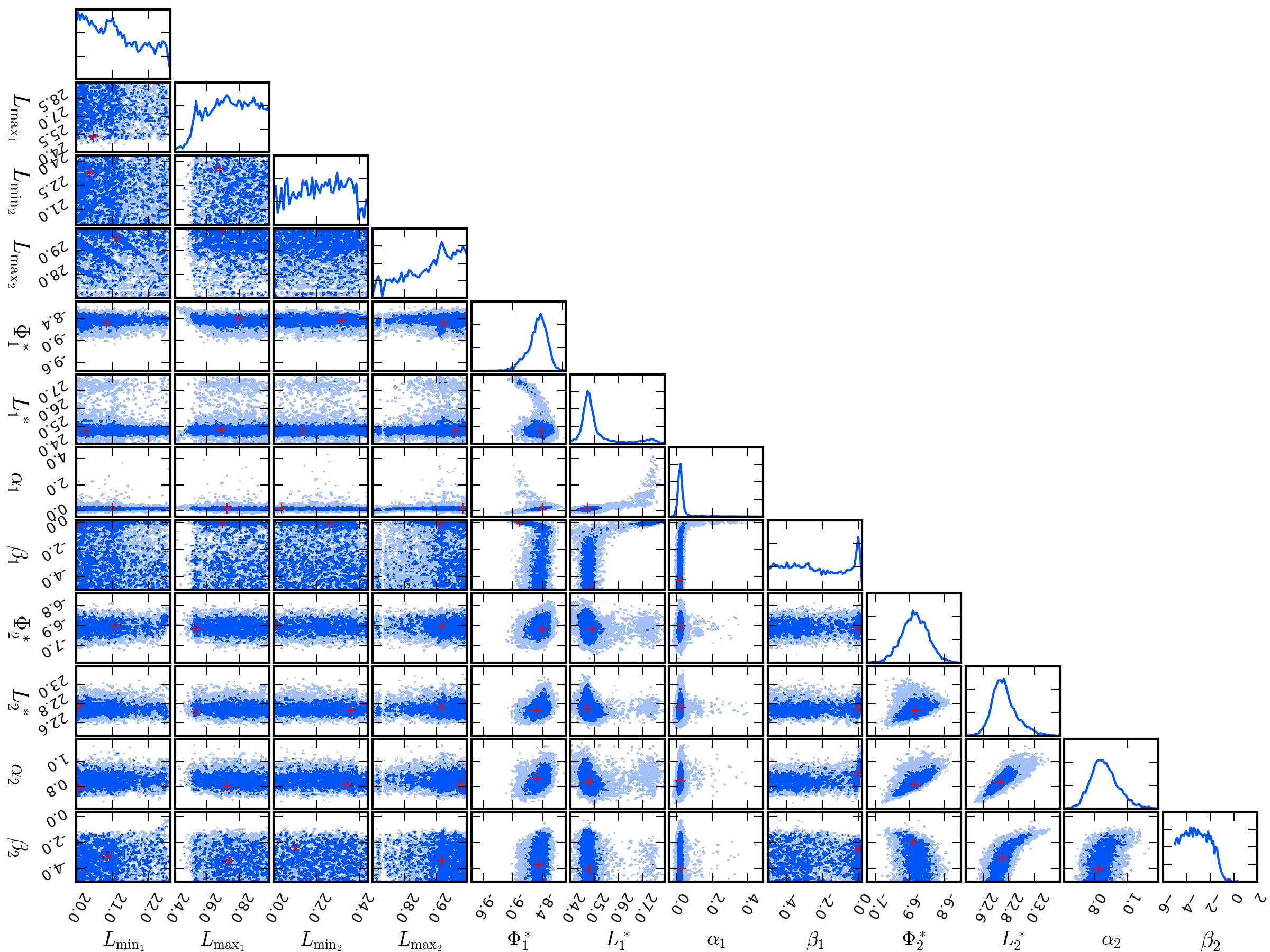}
\end{sideways}
\caption{The posterior distributions of the winning models (\T{table:evidence}) for each redshift bin. The inner plots are the 2-D posterior distributions for the various combinations of the parameters, and the last plots on each row give the 1-D marginalized probability distribution for each parameter. The parameters $L_{\rm{min}_{1,2}}$, $L_{\rm{max}_{1,2}}$, $L_{1,2}*$ and $\Phi_{1,2}^*$ are presented in logarithmic space ($\log_{10}$). The dark blue and the light blue regions are the 68 per cent and 95 per cent regions. The red cross represents the point with the highest marginalised likelihood in each individual plot. The above is for the $0.45 < z < 0.70$ bin.}
\label{fig:tri_1} 
\end{figure*}

\begin{figure*}
\centering
\ContinuedFloat
\begin{sideways}
\includegraphics[width=0.95\textheight]{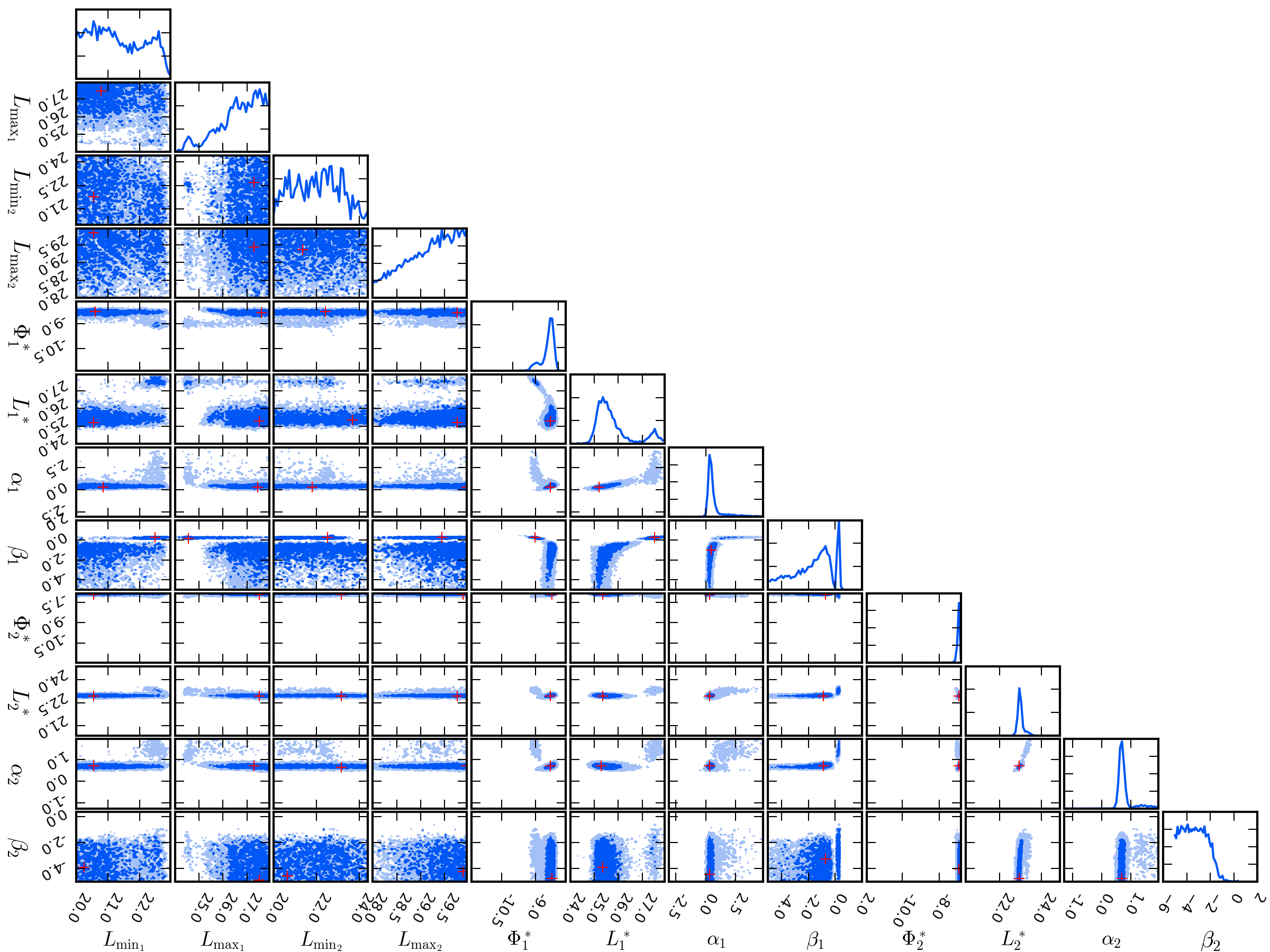}
\end{sideways}
\caption{The posterior distributions, continued. The above is for the $0.70 < z < 1.00$ bin.}
\end{figure*}

\begin{figure*}
\centering
\ContinuedFloat
\begin{sideways}
\includegraphics[width=0.95\textheight]{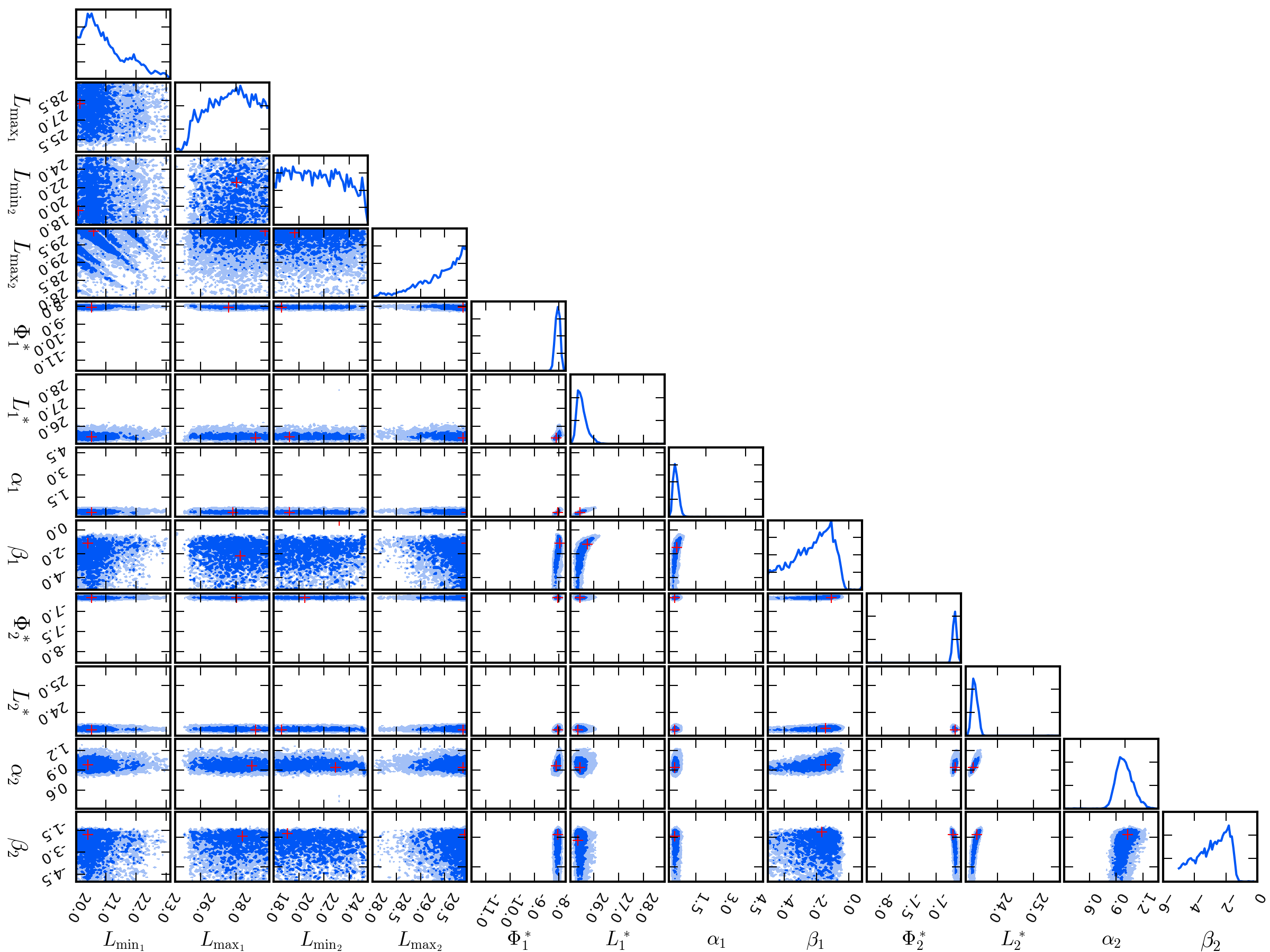}
\end{sideways}
\caption{The posterior distributions, continued. The above is for the $1.00 < z < 1.30$ bin.}
\end{figure*}

\begin{figure*}
\centering
\ContinuedFloat
\begin{sideways}
\includegraphics[width=0.95\textheight]{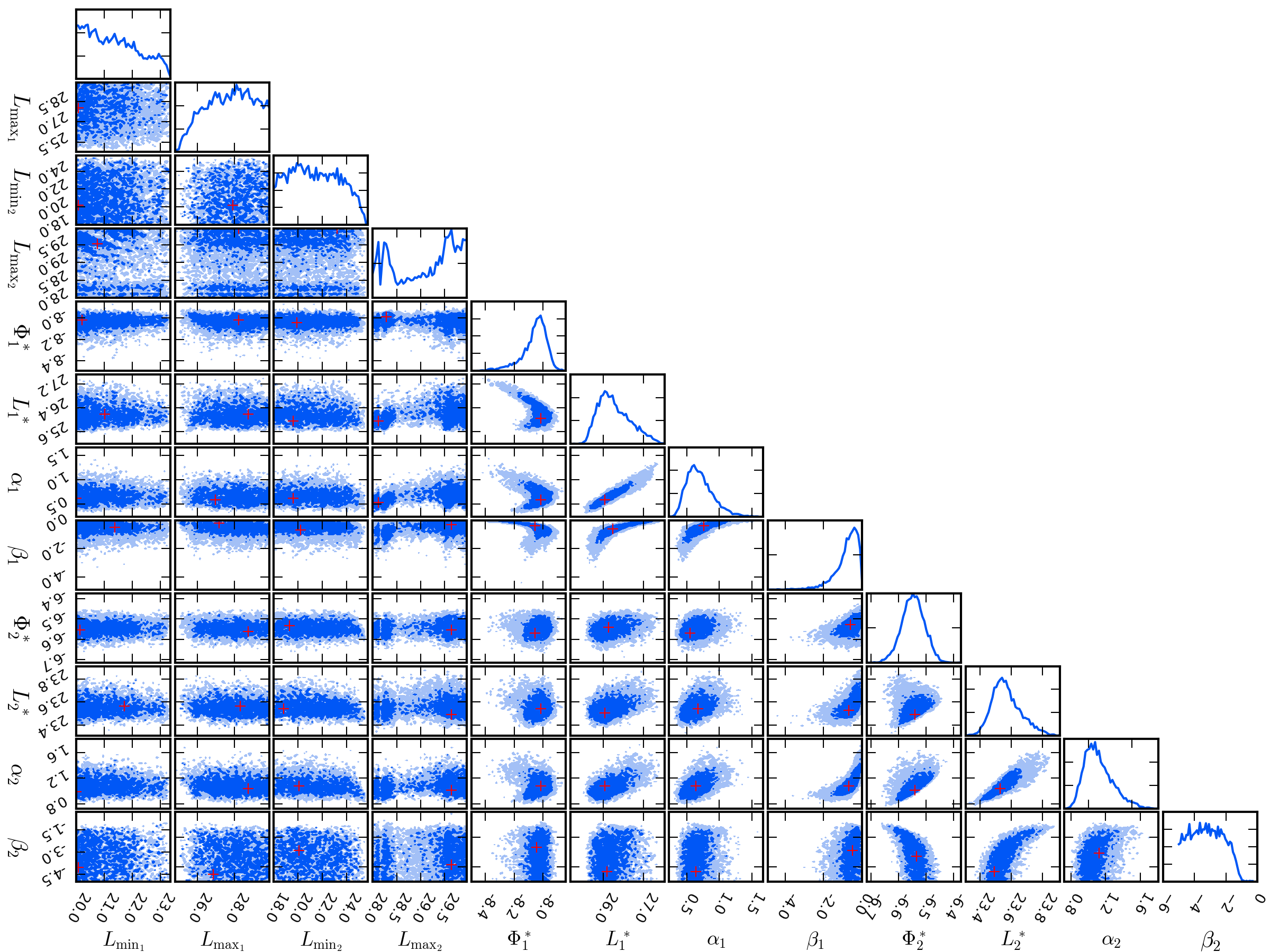}
\end{sideways}
\caption{The posterior distributions, continued. The above is for the $1.30 < z < 1.60$ bin.}
\end{figure*}

\begin{figure*}
\centering
\ContinuedFloat
\begin{sideways}
\includegraphics[width=0.95\textheight]{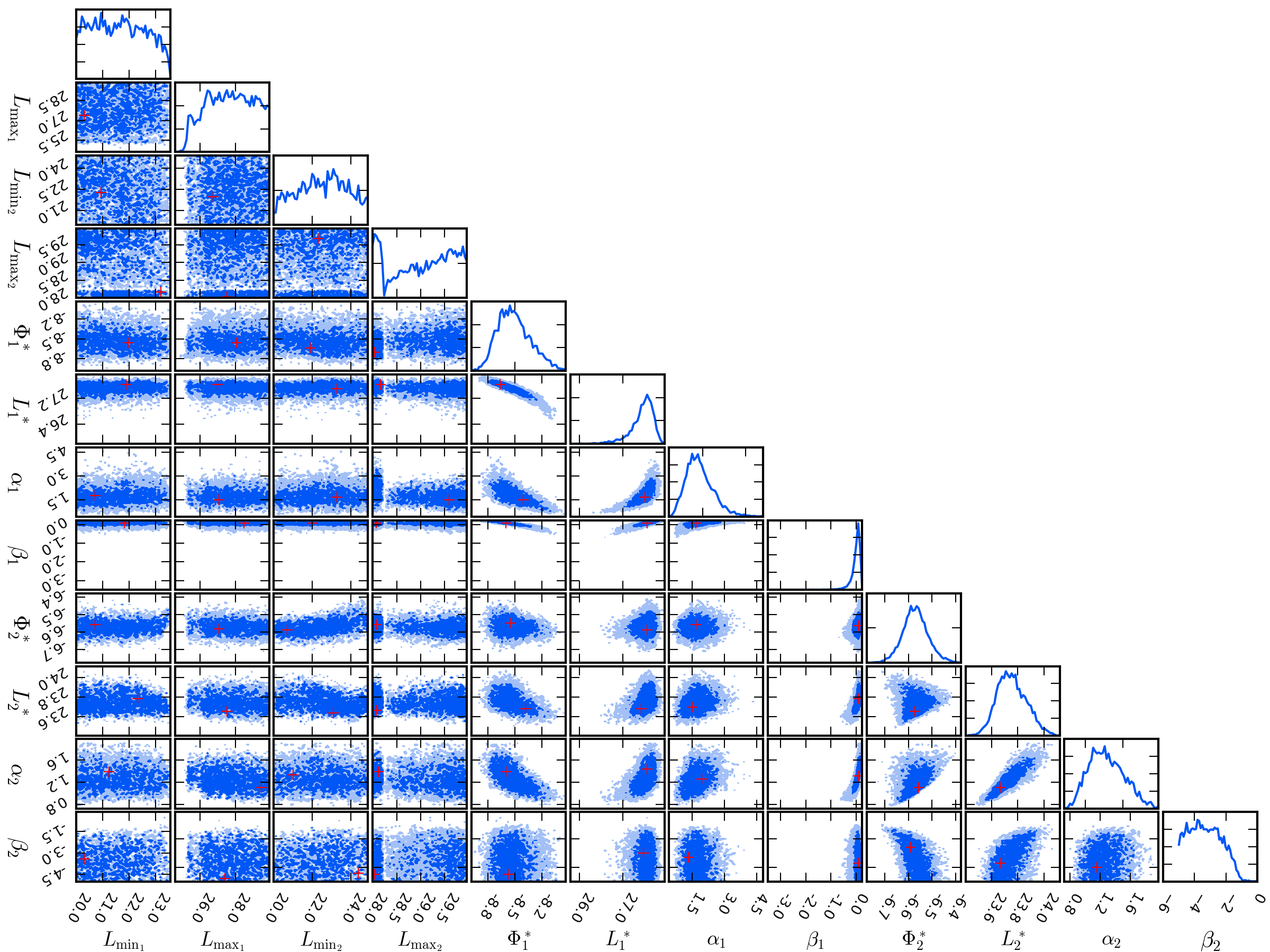}
\end{sideways}
\caption{The posterior distributions, continued. The above is for the $1.60 < z < 1.85$ bin.}
\end{figure*}

\begin{figure*}
\centering
\ContinuedFloat
\begin{sideways}
\includegraphics[width=0.95\textheight]{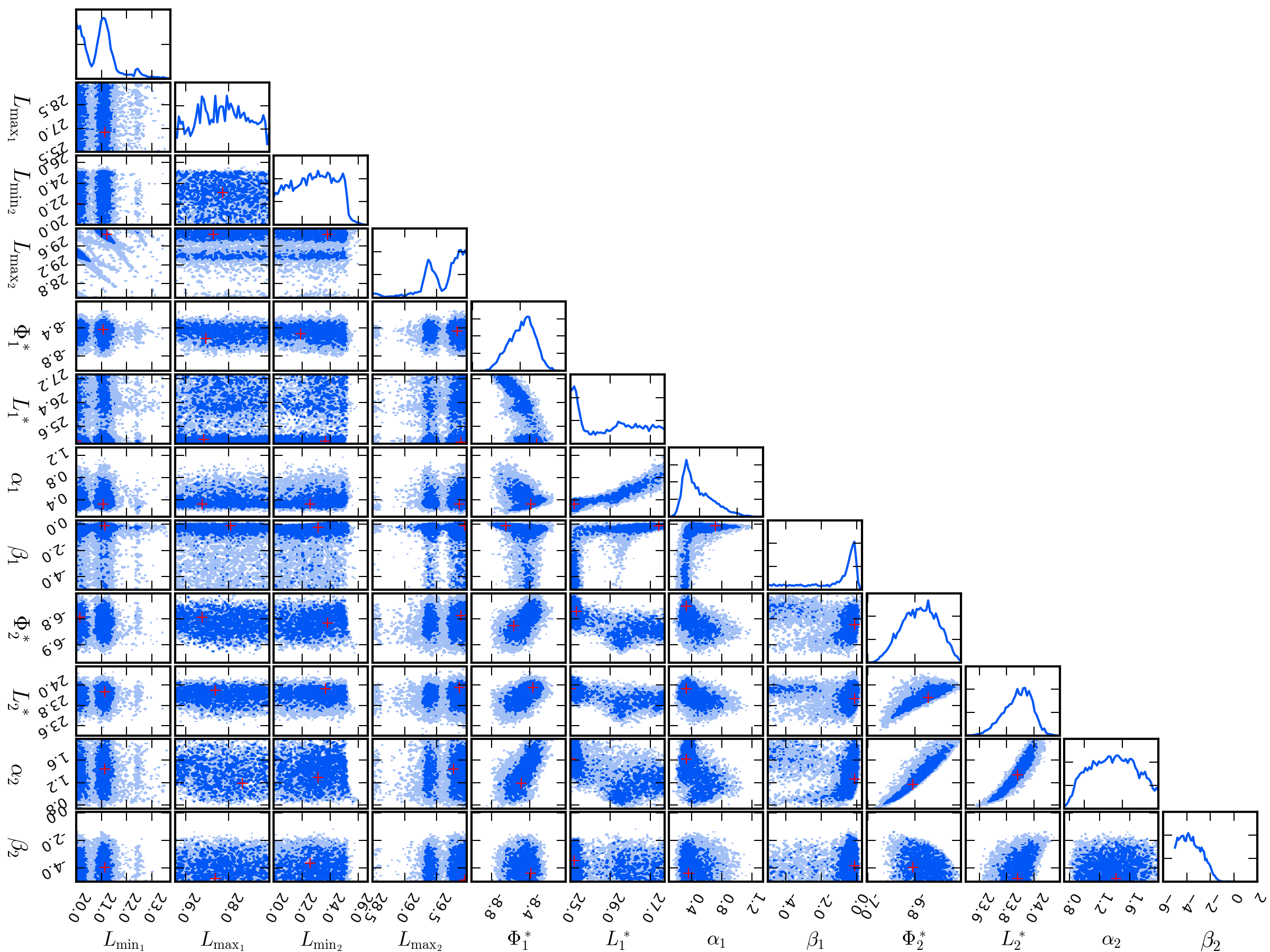}
\end{sideways}
\caption{The posterior distributions, continued. The above is for the $1.85 < z < 2.15$ bin.}
\end{figure*}

\end{appendix}

\label{Bibliography}

\begin{tiny}
\bibliographystyle{mnras}
{\footnotesize 
\bibliography{lit}
}
\end{tiny}

\bsp

\label{lastpage}

\end{document}